\documentclass[lettersize,journal]{IEEEtran}
\usepackage{amsmath,amsfonts,amssymb}
\usepackage{array}
\usepackage{tabularx}
\usepackage[caption=false,font=normalsize,labelfont=sf,textfont=sf]{subfig}
\usepackage{textcomp}
\usepackage{url}
\usepackage{verbatim}
\usepackage{graphicx}
\usepackage{cite}
\usepackage{booktabs}
\usepackage{xcolor}
\usepackage{placeins}
\usepackage{float}

\begin{document}

\makeatletter
\def\@maketitle{%
  \noindent{\footnotesize This work has been submitted to the IEEE for possible publication. Copyright may be transferred without notice, after which this version may no longer be accessible.}
  
  \begingroup
  \centering
  \vskip0.5em%
  {\Huge\@title\par}%
  \vskip1.0em%
  \@author
  \par
  \endgroup
  \vskip1em
}
\makeatother

\bstctlcite{IEEEexample:BSTcontrol}

\title{PilotWiMAE: Pilot-Native Representation Learning for Wireless Channels}

\author{
    \IEEEauthorblockN{
        Berkay Guler, Giovanni Geraci, and Hamid Jafarkhani
        }
\thanks{B.~Guler and H.~Jafarkhani are with the Center for Pervasive Communications and Computing, University of California, Irvine CA, USA. They were supported  in part by the NSF Award CNS-2229467.}
\thanks{G.~Geraci is with Nokia and Universitat Pompeu Fabra, Spain. He was supported in part by grants PID2021-123999OB-I00, PID2024-156488OB-I00, CEX2021-001195-M, and CNS2023-145384.}
\thanks{Part of the results in this paper have been submitted to the International Conference on Machine Learning (ICML), AI4NextG Workshop which is non-archival and will not have a proceedings~\cite{GulerICML2026}.}
}

\maketitle

\begin{abstract}
Channel foundation models assume access to fully observed channels, an assumption that fails in deployment. We introduce PilotWiMAE, a self-supervised framework whose encoder ingests noisy pilot observations directly and whose attention factorizes along the axis separating temporal from joint space-frequency processing, an inductive bias inspired by the physics of the problem. Pilot input shrinks the observation space by up to two orders of magnitude and also removes the unrealistic assumption of full-CSI availability while incurring lower latency. The factorized design generates robust representations by exploiting the separable channel structure and allows a pretraining mask ratio of $99\%$. We pair patch-normalized reconstruction, which captures small-scale fading structure, with an auxiliary scale loss that recovers the large-scale fading features, and use an AWGN curriculum to match pilot noise at pretraining and deployment. Pretrained solely on $3.5$\,GHz and evaluated at $28$\,GHz across in-distribution and out-of-distribution settings, PilotWiMAE's cross-frequency beam selection and channel characterization beat supervised baselines despite operating on a smaller observation space. To weaken the coupling between decoder capacity and representation quality, we further propose a decoder-centric pretraining stage following the encoder-decoder joint pretraining, which allows PilotWiMAE to demonstrate competitive channel estimation without sacrificing representation quality. To foster further work in this direction, we release the PilotWiMAE pretrained weights and training pipeline, together with CSIGen, our Sionna-based ray-tracing channel-generation tool, and the channel datasets used in this work.
\end{abstract}

\begin{IEEEkeywords}
Self-supervised learning, wireless channel representation, foundation models, AI-native wireless systems
\end{IEEEkeywords}

\section{Introduction}
\label{sec:intro}

\IEEEPARstart{R}{ecent} channel foundation models have made substantial progress in learning transferable representations of wireless channels by pretraining and evaluating fully observed channels generated by stochastic or ray-tracing simulators \cite{AlikhaniLWM2025, Jiang2025, Jiang2026, LiuWiFo2025, AlikhaniLWMTemporal2026, Yang2026, LiuLLM4CP2024, Guler2026, LiuWiFo-CF2025, Wang2026, Sheng2025, Pan2025, Zhou2026, LiuWiFo22026, Aboulfotouh2025, Guo2025, Zheng2026}. Some of these works add i.i.d.\ additive white Gaussian noise (AWGN) to fully observed channels as a concession to realism \cite{Jiang2026, Guler2026, LiuWiFo-CF2025, LiuWiFo2025, LiuLLM4CP2024, AlikhaniLWM2025, Yang2026, AlikhaniLWMTemporal2026, LiuWiFo22026, Guo2025, Zheng2026, Aboulfotouh2025}, while others omit noise evaluation altogether. Neither of these two cases reflects how errors in channel state information (CSI) arise in practice. In a real receiver, the channel is estimated from pilots, and only the error at pilot resource elements is i.i.d.\ AWGN (when there is no interference from pilot reuse) \cite{Edfors1998}. The error at non-pilot resource elements, which make up the vast majority of the grid, depends on the interpolation method, the channel's delay-Doppler structure, the pilot density, the SNR at the pilots, and the pilot design, and does not admit a simple i.i.d.\ model \cite{Coleri2002}. Evaluation under fully observed or i.i.d.-perturbed channels therefore can only characterize how well a model learns the channel structure, but leaves open how it behaves in a system where such channels are never available. Given the known sensitivity of the learned methods to noise and distribution shift \cite{Hendrycks2019, Taori2020}, this gap is worth closing.

The second gap concerns the cost of deployment. Wireless foundation models have largely inherited transformer architectures \cite{Vaswani2017, Dosovitskiy2021} and training recipes \cite{Devlin2019, He2022} from vision and language, where parameter count and sequence length do not face a hard runtime ceiling, while performance improves predictably with additional data and computation \cite{Kaplan2020, Hoffmann2022, Zhai2022}. In wireless systems, tasks such as precoding, scheduling, and decoding must be completed within slot-level timing budgets of the order of a millisecond or less \cite{3GPP_TS38.211}. However, computational footprint is rarely reported in the literature on channel foundation models. When reported, the results often rely on high-end GPUs and non-uniform optimization stacks (e.g., quantization or FlashAttention \cite{Dao2022}), making even large models appear practical and masking true deployment cost.

\begin{figure*}[!t]
  \centering
  \includegraphics[width=\textwidth]{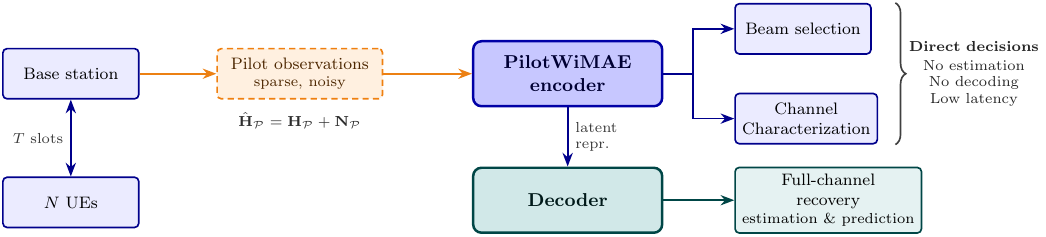}
  \caption{High-level PilotWiMAE pipeline: The model consumes sparse noisy pilot observations directly, pilot representations support direct decision-making tasks without channel estimation or decoding, while an optional decoder reconstructs the full channel for estimation and/or prediction.}
  \label{fig:intro-overview}
\end{figure*}

We address both gaps with two co-equal design principles. First, we pursue \emph{robustness by design} by operating directly on sparse, noisy pilot observations. Our approach removes an explicit channel estimator from the critical path to prevent error propagation at realistic low SNR, matches deployment observables, eliminates the full-CSI assumption of prior channel foundation models, and cleanly integrates with existing pilot-based protocols. Channel recovery is posed as a downstream task on the same learned representations and is naturally aligned with our reconstruction-based pretraining objective.

Second, we enforce \emph{wireless specificity by design} by factorizing attention along temporal and space-frequency domains, an inductive bias grounded in the wide-sense stationary uncorrelated scattering (WSSUS) model \cite{Bello1963} and its MIMO extension \cite{Matz2011}, where temporal and spectro-spatial correlations arise from distinct physical mechanisms. The same principle motivates our pretraining objective, which pairs patch-normalized reconstruction for small-scale fading with an auxiliary scale loss that recovers large-scale fading statistics. Overall, pilot input shrinks the observation space, while the factorized design exploits separable channel structure to support an aggressive 99\% pretraining mask ratio. Together, they yield sub-millisecond inference latency and representations that remain reliable in the noisy, partially observed regime where decisions are actually made.

Fig.~\ref{fig:intro-overview} provides a high-level summary of this deployment-oriented pipeline. We next situate PilotWiMAE with respect to recent work.

\subsection{Related work}

Prior work on wireless channel foundation models broadly adopts (i) joint-embedding methods, including contrastive learning, \cite{Jiang2025, Guler2026, Pan2025, Salihu2024, Chu2026}, (ii) masked reconstructive learning \cite{Jiang2026, Guler2026, AlikhaniLWM2025, AlikhaniLWMTemporal2026, Yang2026, LiuWiFo22026, Pan2025, Aboulfotouh2025, LiuWiFo2025}, and (iii) joint reconstructive-contrastive objectives \cite{Pan2025, Guler2026}. Alongside these encoder-oriented families, a separate line adopts decoder-only causal generation for temporal prediction and forecasting \cite{Sheng2025, LiuLLM4CP2024, Zhou2026}. Although effective for sequence generation, decoder-only models are not representation learners in the encoder-based sense because they do not directly expose a compact, task-agnostic representation for downstream adaptation. We therefore focus on encoder-based self-supervised pipelines.

This focus still leaves important design choices. Contrastive objectives can learn transferable wireless features, but they often require multiple augmented views and forward passes before each update, and their performance is highly dependent on view construction and positives/negatives \cite{Jiang2025, Guler2026, Pan2025}. Non-contrastive joint-embedding methods, such as JEPA-style predictors over masked latents \cite{Chu2026, Salihu2024}, avoid explicit negative pairs and typically rely on masked-context prediction consistency rather than handcrafted augmentation pipelines, but for the same reason they do not explicitly learn reconstruction-oriented features. However, in wireless, channel reconstruction tasks are first-class downstream objectives. It is desirable that the pretraining objective shapes representations that retain the signal structure needed to recover the channel itself, not just abstract latent invariances. Within reconstructive learning, BERT-style \cite{Devlin2019} masked modeling feeds a dense sequence to the encoder, processes masked and visible tokens together, and uses lightweight heads to only predict the masked positions \cite{AlikhaniLWM2025, Yang2026, Pan2025, AlikhaniLWMTemporal2026}. However, this paradigm scales poorly with mask ratio because the encoder still processes masked tokens that carry no observation content.

MAE-style pretraining bypasses these limitations. The encoder processes only visible tokens, and a transformer decoder reconstructs the masked content from encoded visible tokens and mask placeholders \cite{He2022}. The encoder cost decreases with the mask ratio, single-view single-pass updates avoid the multi-view overhead of contrastive pretraining, and the input-space reconstruction objective preserves channel-valued structure in the learned representation. The theory further shows that masked reconstruction implicitly performs contrastive alignment, because different masked views of the same input that share a reconstruction target act as positive pairs and are pulled together in feature space. This explains the quality of MAE's representation without an explicit contrastive loss \cite{Zhang2022}.

Nevertheless, two design choices shape what an MAE encoder actually learns. The first choice is the reconstruction target. All wireless MAE variants use raw MSE reconstruction \cite{Jiang2026, Guler2026, LiuWiFo22026, Aboulfotouh2025, LiuWiFo2025}, which is poorly matched to channels whose amplitudes span a very large dynamic range. Under raw MSE, the loss is dominated by a small fraction of high-power, often LoS-like channels, while low-power NLoS-rich channels, whose small-scale fading patterns carry the complex multipath structure the encoder should actually learn, contribute negligibly to the gradient. The second choice is how representational work is split between encoder and decoder. Vision MAE pairs a deep encoder with a shallow decoder and discards the decoder after pretraining, forcing representational load onto the encoder \cite{He2022}. Wireless follows this trend. However, the decoder is also maintained and reused for channel reconstruction tasks (channel estimation, prediction, and CSI feedback) 
\cite{Jiang2026, Guler2026, LiuWiFo22026, LiuWiFo2025}, since both strong representation quality and accurate reconstruction are desired. As a result, a single pretraining stage is forced to deliver both objectives at once, making the encoder-decoder capacity split a compromise rather than a deliberate choice. PilotWiMAE addresses both issues as explained in detail in Section~\ref{sec:approach}. 

The choice of objective and reconstruction loss is only part of what makes a pretrained channel representation useful in practice. Two further dimensions matter just as much. The first is \emph{the input interface}, which determines what the encoder actually observes during pretraining and deployment. The second, \emph{the architectural inductive bias}, regulates which channel properties the encoder is structurally encouraged to exploit, with direct consequences for representation quality. Both remain underexplored in wireless self-supervised learning, and we discuss them in turn.

Starting with the input interface, most existing protocols still assume full-grid CSI at evaluation, and operate on full-grid CSI tensors during pretraining, sometimes masked under a self-supervised reconstruction objective. Several works are entirely based on pretraining and evaluating clean full-CSI \cite{AlikhaniLWM2025, Jiang2025, Jiang2026, LiuWiFo-CF2025}. Others inject i.i.d.\ AWGN into channels during pretraining and/or evaluation \cite{LiuWiFo2025, Guler2026, Yang2026, LiuLLM4CP2024, LiuWiFo22026}, which improves stress testing, but still does not match pilot-based observability in real receivers. Even when sparse settings are considered, they are typically task-specific (e.g., localization under pilot-position sampling) or use clean sparse observations rather than noisy pilot measurements \cite{Pan2025}. Consequently, a key mismatch in the deployment remains. In these systems, representations are largely learned and validated in regimes where dense CSI is available, whereas practical systems observe noisy CSI only at pilot resource elements, without the possibility of recovering the perfect CSI.

A small number of studies are moving toward realistic receiver-side conditions. \cite{Wang2026} prepends a channel estimation and refinement module to a frozen feature extraction network. Architecturally, this is not different from cascading any channel estimator with a feature extractor and reinstates the full-CSI assumption at the encoder input rather than establishing a pilot-native, general-purpose representation learner across tasks. \cite{Zhou2026} pretrains a causal autoregressive model on historical LMMSE channel estimates rather than on clean full-CSI, and at inference uses the autoregressively predicted state as a prior for refining the channel estimate from current pilots. The focus is therefore generative forecasting and prior-based channel estimation, rather than self-supervised encoder representation learning across tasks.

Another challenge concerns architectural inductive bias. Most prior channel representation models apply dense all-to-all attention to channel tokens, even though channel statistics exhibit structure induced by the underlying physics. A recent angle-delay-time representation learner sparsifies attention to tokens falling inside an angle-delay window and a temporal cone, capturing scatterer evolution in a sparse angle-delay domain \cite{AlikhaniLWMTemporal2026}. That construction does not carry over to dense space-frequency inputs, where correlations are not confined to a small neighborhood of each token, so spectro-spatial mixing must attend broadly over the space-frequency grid rather than only locally. Even so, a mild WSSUS assumption implies that temporal correlation factorizes from joint space-frequency correlation \cite{Bello1963, Matz2011}, which motivates factorized attention. Factorized attention appears in video \cite{Bertasius2021, Arnab2021} and in concurrent wireless work aimed at complexity reduction \cite{Yellapragada2026, Zubow2026, Yang2026}, but not as a pilot-native, physics-induced bias to separate temporal from spectro-spatial domains for self-supervised CSI representation learning. PilotWiMAE addresses this gap by jointly aligning \emph{what is observed} (noisy pilots) and \emph{how it is processed} (WSSUS-motivated factorized attention) in one pretraining framework.

\subsection{Contribution and Summary of Results}

We introduce PilotWiMAE\footnote{PilotWiMAE code, pretrained weights, and training pipeline are available at \url{https://github.com/BerkIGuler/PilotWiMAE}. CSIGen, our Python-based ray-tracing channel-generation tool built on Sionna, and channel datasets used in this work are available at \url{https://github.com/BerkIGuler/CSIGen}.}, a self-supervised, foundation-model-style framework for wireless channel representation. The key contributions and findings of this work are summarized below.
\begin{itemize}
    \item We propose a pilot-native input interface that operates directly on noisy pilot observations for decision-making tasks, bypasses channel estimation, and eliminates the need for the full-CSI assumption typically made by prior channel foundation models.
    \item We design a factorized space-time (FST) encoder that applies temporal attention across time slots and spectro-spatial attention within each time slot, an inductive bias grounded in WSSUS-motivated separability that yields robust, high-performing representations and enables an aggressive $99\%$ pretraining mask ratio.
    \item We pair patch-normalized reconstruction, which captures small-scale fading structure, with an auxiliary scale loss that recovers the large-scale fading signature (path loss and shadowing), jointly supervised on encoder- and decoder-side features.
    \item We introduce a decoder-centric masked reconstruction pretraining stage, following the AWGN-augmented joint encoder-decoder pretraining with masked reconstruction and scale losses, to decouple decoder capacity from encoder representation quality.
    \item Through extensive evaluation under cross-band ($3.5$ to $28$\,GHz) transfer, including both in-distribution (ID) and out-of-distribution (OOD) settings, we demonstrate that PilotWiMAE's frozen representations transfer to beam selection and channel characterization without task-specific fine-tuning, beating supervised baselines despite operating on a smaller observation space.
    \item We further demonstrate noise-robust, pilot-pattern-agnostic channel estimation that remains competitive with supervised pilot-pattern-specific baselines, and report how performance scales with decoder depth when only the decoder is pretrained on top of a frozen encoder.
    \item We introduce a noise-robust AWGN pretraining curriculum that anneals the SNR lower bound across epochs, progressively exposing the model to more challenging SNR conditions while aligning corruption with pilot-noise statistics.
\end{itemize}

\begin{figure*}[t]
  \centering
  \includegraphics[width=\textwidth]{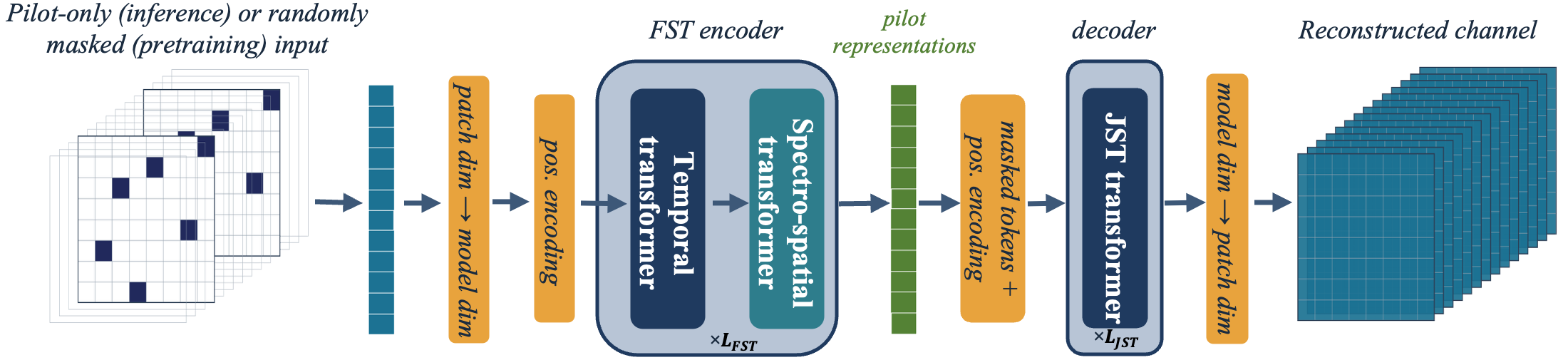}
  \caption{PilotWiMAE architecture. Pilot patches feed the FST encoder. The resulting representations are decoded by a JST transformer. Finally, the tokens are mapped back to patch dimension for channel reconstruction. For clarity, the diagram shows the main forward path only and omits several elements described in Section~\ref{sec:approach}.}
  \label{fig:architecture}
\end{figure*}

\subsubsection*{Paper organization}
The remainder of this paper is organized as follows. Section~\ref{sec:problem} formulates the pilot-native channel representation problem. Section~\ref{sec:approach} details the PilotWiMAE framework (Fig.~\ref{fig:architecture}). Section~\ref{sec:experiments} describes the experimental setup and reports cross-band ID and OOD results. Section~\ref{sec:complexity} reports training and inference complexity. Section~\ref{sec:conclusion} concludes the paper.

\section{Problem Formulation}
\label{sec:problem}

We consider a base station operating over $T$ consecutive time slots, each with $N_{\mathrm{sc}}$ subcarriers and a uniform planar array of $N_{\mathrm{h}} \times N_{\mathrm{v}}$ antennas. The full channel tensor $\mathbf{H} \in \mathbb{C}^{T \times N_{\mathrm{h}} N_{\mathrm{v}} \times N_{\mathrm{sc}}}$ is never observed. Instead, the receiver obtains a noisy observation at a sparse set of pilot resource elements indexed by $\mathcal{P}$,
\begin{equation}
    \hat{\mathbf{H}}_\mathcal{P} = \mathbf{H}_\mathcal{P} + \mathbf{N}_\mathcal{P}(\mathrm{SNR}),
    \label{eq:pilot-obs}
\end{equation}
where $\mathbf{N}_\mathcal{P}(\mathrm{SNR})$ is i.i.d.\ circularly-symmetric complex Gaussian noise whose variance is a known function of the pilot SNR alone \cite{Edfors1998}.

More precisely, pilots are sent at selected OFDM symbol indices $\mathcal{T}_{\mathrm{p}} \subset \{0, \ldots, T-1\}$ and subcarrier indices $\mathcal{F}_{\mathrm{p}} \subset \{0, \ldots, N_{\mathrm{sc}}-1\}$, and at each $(t, f) \in \mathcal{T}_{\mathrm{p}} \times \mathcal{F}_{\mathrm{p}}$ all $N_{\mathrm{h}} N_{\mathrm{v}}$ antenna entries are observed. Our method is agnostic to the specific layout of $(\mathcal{T}_{\mathrm{p}}, \mathcal{F}_{\mathrm{p}})$.

Prior channel foundation models largely adopt a dense CSI interface rather than the pilot-only observations in~\eqref{eq:pilot-obs}. A generic way to express the mismatch between such a full-grid tensor and ground truth $\mathbf{H}$ is an additive estimation residual,
\begin{equation}
  \hat{\mathbf{H}} = \mathbf{H} + \mathbf{E}(\boldsymbol{\theta}),
\end{equation}
where $\mathbf{E}(\boldsymbol{\theta})$ is generally correlated and its statistics depend on the estimator and on channel parameters that are scenario-specific and unavailable in general. Existing pipelines either ignore $\mathbf{E}(\boldsymbol{\theta})$ or substitute it with i.i.d.\ circularly-symmetric complex Gaussian noise. While mathematically convenient, the true residual is not i.i.d.\ circularly-symmetric complex Gaussian.

One straightforward alternative is to add a channel estimator to the encoder, recovering $\hat{\mathbf{H}}$ before producing representations. This either requires end-to-end training of an additional block or requires the encoder to be robust to the artifacts of a specific estimator, both of which complicate the design without addressing the root issue. Instead, we feed $\hat{\mathbf{H}}_\mathcal{P}$ directly into the encoder, i.e., 
\begin{equation}
  \hat{\mathbf{H}}_\mathcal{P} \;\longrightarrow\; \text{encoder} \;\longrightarrow\; \text{representations} \;\longrightarrow\; \text{decision}.
\end{equation}

The hypothesis is that when pretraining uses masked reconstruction, sparse pilot inputs can support the global channel structure in the encoder latent. A deliberately shallow decoder turns the encoder latent into the representational bottleneck where the encoder must integrate information across the resource grid from the visible pilot tokens to predict the masked content. Decision-making tasks that do not require dense CSI, including beam selection and channel characterization, can follow a compact inference path using pilot-only features without reconstructing the entire grid during inference. In general, joint encoder-decoder pretraining couples latent quality with decoder capacity, motivating us to discard that decoder, freeze the encoder, and pretrain a freshly initialized decoder with greater reconstruction capacity on patch-normalized reconstruction alone. The second stage yields stronger dense-channel recovery while encoder representations stay fixed.

\section{The PilotWiMAE Framework}
\label{sec:approach}

In this section, we introduce PilotWiMAE, a self-supervised framework that realizes the two design principles of Section~\ref{sec:intro} through six components, each developed in its own subsection below.

\subsection{Pilot-native input interface}
During pretraining, the encoder sees extremely sparse inputs generated by structured random masking under AWGN. During inference, it sees a sparse input $\hat{\mathbf{H}}_\mathcal{P}$ induced by the pilot pattern (Appendix~\ref{app:pilot-pattern}).

Before tokenization, each sample is power-normalized using a dataset-level reference power $P_{\mathrm{ref}}$ computed on the pretraining split. We use $\bar{\mathbf{H}}=\mathbf{H}/\sqrt{P_{\mathrm{ref}}}$. Defining $S=N_{\mathrm{h}}N_{\mathrm{v}}$ and $F=N_{\mathrm{sc}}$, let the complex input tensor be $\bar{\mathbf{H}}\in\mathbb{C}^{T\times S\times F}$. Let the 3D patch size be $(p_{\mathrm{t}},p_{\mathrm{s}},p_{\mathrm{f}})$. Defining $(n_{\mathrm{t}},n_{\mathrm{s}},n_{\mathrm{f}})=(T/p_{\mathrm{t}},S/p_{\mathrm{s}},F/p_{\mathrm{f}})$, tokenization yields $P=n_{\mathrm{t}} n_{\mathrm{s}} n_{\mathrm{f}}$ patches, with $N_{\mathrm{sf}}=n_{\mathrm{s}} n_{\mathrm{f}}$ spectro-spatial tokens per slot. The real and imaginary parts are split and concatenated within each patch. Therefore, each raw patch vector has dimension $D_{\mathrm{p}}=2p_{\mathrm{t}} p_{\mathrm{s}} p_{\mathrm{f}}$ before linear projection to the model dimension $d$.

Patch offsets $(i_{\mathrm{t}},i_{\mathrm{s}},i_{\mathrm{f}})$ index temporal, spatial, and frequency patches with $i_{\mathrm{t}}\in\{0,\ldots,n_{\mathrm{t}}-1\}$, $i_{\mathrm{s}}\in\{0,\ldots,n_{\mathrm{s}}-1\}$, and $i_{\mathrm{f}}\in\{0,\ldots,n_{\mathrm{f}}-1\}$, respectively, and time-major unfolding maps them to $p=i_{\mathrm{t}} n_{\mathrm{s}} n_{\mathrm{f}}+i_{\mathrm{s}} n_{\mathrm{f}}+i_{\mathrm{f}}\in\{0,\ldots,P-1\}$. Each token $\mathbf{x}_p \in \mathbb{R}^{d}$ is the linear projection of the $D_{\mathrm{p}}$-dimensional real-imaginary channel patch at flat index $p$ to the model dimension.

We add an axial sinusoidal positional embedding to each token. To this end, we partition the model dimension into per-axis sub-widths $d=d_{\mathrm{t}}+d_{\mathrm{s}}+d_{\mathrm{f}}$ with $d_{\mathrm{t}}=d_{\mathrm{s}}=\lfloor d/3\rfloor$ and $d_{\mathrm{f}}=d-2\lfloor d/3\rfloor$, and assign one sub-width to each axis. For axis $a \in \{\mathrm{t},\mathrm{s},\mathrm{f}\}$ with patch count $n_a$ and sub-width $d_a$, the per-axis embedding $\boldsymbol{\psi}^{(a)}_{i} \in \mathbb{R}^{d_a}$ at index $i \in \{0,\ldots,n_a-1\}$ uses the standard $1$D transformer sinusoidal recipe of \cite{Vaswani2017},
\begin{align}
  [\boldsymbol{\psi}^{(a)}_{i}]_{2 j} &= \sin\!\bigl(i\,10\,000^{-2j/d_a}\bigr),\label{eq:pe-sin}\\[-2pt]
  [\boldsymbol{\psi}^{(a)}_{i}]_{2 j+1} &= \cos\!\bigl(i\,10\,000^{-2j/d_a}\bigr),\label{eq:pe-cos}
\end{align}
with $j \in \{0,\ldots,\lfloor (d_a-1)/2 \rfloor\}$, applying \eqref{eq:pe-cos} only when $2j+1<d_a$. The patch-level positional embedding then concatenates the three axis blocks, $\mathbf{e}^{\mathrm{pe}}_{p} = \bigl[\,(\boldsymbol{\psi}^{(\mathrm{t})}_{i_{\mathrm{t}}})^{\top},\ (\boldsymbol{\psi}^{(\mathrm{s})}_{i_{\mathrm{s}}})^{\top},\ (\boldsymbol{\psi}^{(\mathrm{f})}_{i_{\mathrm{f}}})^{\top}\,\bigr]^{\top}\in\mathbb{R}^{d}$, with $(i_{\mathrm{t}},i_{\mathrm{s}},i_{\mathrm{f}})$ decoded from $p$. Tokens receive this positional encoding via $\mathbf{x}_{p}\leftarrow \mathbf{x}_{p}+\alpha_{\mathrm{pe}}\mathbf{e}^{\mathrm{pe}}_{p}$, where $\alpha_{\mathrm{pe}}$ is a learnable scalar initialized near zero such that the fixed sinusoidal magnitudes do not dominate the linear patch embedding early in pretraining.

\subsection{WSSUS-motivated factorized attention}
Under the classical WSSUS model \cite{Bello1963} and its MIMO extension \cite{Matz2011}, the temporal correlation is governed by the Doppler spectrum, a function of the environment mobility, while the spectro-spatial correlation is governed by the joint angular-delay power spectrum, a function of the scattering geometry. Writing the channel autocorrelation over a time lag $\Delta t$, a frequency lag $\Delta f$, and a spatial lag $\Delta s$, the WSSUS assumption and the standard separability of Doppler from angle-delay dispersion in MIMO-WSSUS channels \cite{Matz2011} yield
\begin{equation}
  R_{\mathrm{H}}(\Delta t, \Delta f, \Delta s) \;\approx\; R_{\mathrm{t}}(\Delta t)\,R_{\mathrm{sf}}(\Delta f, \Delta s).
  \label{eq:wssus-factorization}
\end{equation}
According to \eqref{eq:wssus-factorization}, temporal and space-frequency correlations are weakly coupled and a representation learner does not need to model cross-domain correlations. On the other hand, space and frequency remain coupled through the jointly angle- and delay-dependent scattering structure \cite{Matz2011}. Therefore, a three-way separability is not desirable as it would discard the real physical structure that the encoder should learn.

\begin{figure}[b]
  \centering
  \includegraphics[width=\linewidth]{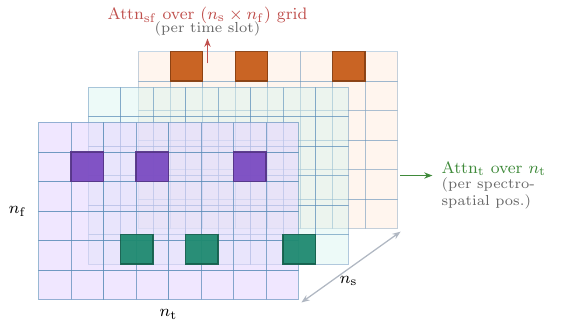}
  \caption{Factorized space-time attention on the patch-token grid with axes $(n_{\mathrm{t}}, n_{\mathrm{s}}, n_{\mathrm{f}})$. For illustration, the figure uses $n_{\mathrm{t}}=8$, $n_{\mathrm{s}}=3$, $n_{\mathrm{f}}=6$, $n_{\mathrm{k}}=3$, and $\rho_{\mathrm{k}}=\frac{3}{18}$, which leads to $\approx 94\%$ mask ratio.}
  \label{fig:fst-attention}
\end{figure}

This prior maps to an attention factorization. To represent the sparse subset of tokens retained by the input mask, let $n_{\mathrm{k}}\le n_{\mathrm{t}}$ and $N'_{\mathrm{sf}}\le N_{\mathrm{sf}}$ denote the numbers of retained temporal and spectro-spatial patch indices, respectively. The retained set is rectangular both during pretraining (structured random masking, Section~\ref{sec:approach:masking}) and at inference (the fixed pilot pattern, Appendix~\ref{app:pilot-pattern}), and $(n_{\mathrm{k}}, N'_{\mathrm{sf}})=(n_{\mathrm{t}}, N_{\mathrm{sf}})$ recovers the unmasked case. Let $\mathbf{Z}^{(\ell)} \in \mathbb{R}^{n_{\mathrm{k}} \times N'_{\mathrm{sf}} \times d}$ be the token tensor at the input to Block $\ell \in \{0,1,\ldots,L_{\mathrm{FST}}\}$, where $L_{\mathrm{FST}}$ is the number of FST encoder blocks. Within each block, our FST encoder applies temporal attention across the retained temporal-patch indices followed by spectro-spatial attention within each retained temporal slice, i.e.,
\begin{align}
  \mathbf{Z}^{(\ell+\tfrac{1}{2})}_{:,s,:} &= \mathrm{Attn}_{\mathrm{t}}\!\left(\mathbf{Z}^{(\ell)}_{:,s,:}\right), \quad s=1,\dots,N'_{\mathrm{sf}}, \label{eq:fst-t}\\
  \mathbf{Z}^{(\ell+1)}_{t,:,:} &= \mathrm{Attn}_{\mathrm{sf}}\!\left(\mathbf{Z}^{(\ell+\tfrac{1}{2})}_{t,:,:}\right), \quad t=1,\dots,n_{\mathrm{k}}, \label{eq:fst-sf}
\end{align}
where $\mathbf{Z}^{(\ell+\tfrac{1}{2})}$ denotes the intermediate token tensor between the two sublayers. Residual connections and feedforward sublayers are omitted for the sake of clarity. As illustrated in Fig.~\ref{fig:fst-attention}, cross-slot information is exchanged only by temporal attention at fixed spectro-spatial indices. On the other hand, spectro-spatial information is exchanged only by within-slot attention at fixed time indices, mirroring \eqref{eq:wssus-factorization}. Because factorization is an inductive bias rather than a sparsification of the attention matrix, it does not degrade expressivity on structures that WSSUS captures. Our ablations against a joint space-time (JST) baseline of matched parameter count show consistent gains on both ID and OOD beam selection and channel characterization.

Beyond the inductive-bias argument, factorization also lowers per-layer attention cost relative to joint space-time attention. We will discuss the complexity expressions, measured FLOPs, and per-sample latency in Section~\ref{sec:complexity}.

\subsection{Aggressive masking enabled by factorization}
\label{sec:approach:masking}
Factorized attention and separability permit an unusually aggressive pretraining masking regime. We apply a structured random mask that retains only $n_{\mathrm{k}}$ out of $n_{\mathrm{t}}$ temporal-patch indices. Across those retained patches, we keep a common fraction $\rho_{\mathrm{k}} \in (0,1)$ of the spectro-spatial token positions (the same positions in every retained patch rather than resampled per patch). Therefore, the overall fraction of visible tokens is $(n_{\mathrm{k}}/n_{\mathrm{t}})\,\rho_{\mathrm{k}}$ and the overall mask ratio is $1 - (n_{\mathrm{k}}/n_{\mathrm{t}})\,\rho_{\mathrm{k}}$. This is equivalent to $N'_{\mathrm{sf}} = \rho_{\mathrm{k}} N_{\mathrm{sf}}$ in ~\eqref{eq:fst-t}. The mask structure matches the factorization. The temporal block mixes information across the $n_{\mathrm{k}}$ kept temporal patches at each fixed spectro-spatial position. The spectro-spatial block mixes across the visible positions within each kept patch. Therefore, the decoder receives informed tokens at every visible location. Randomizing the mask across pretraining examples keeps the encoder agnostic to any specific pilot pattern. As a result, at inference, the same pretrained model can ingest whatever fixed pilot configuration is used at the receiver.

This structured keep set is well matched to factorized attention since every attended pair under FST differs on at most one axis. Under \eqref{eq:wssus-factorization}, the autocorrelation along a single axis is governed by either $R_{\mathrm{t}}$ alone or $R_{\mathrm{sf}}$ alone. Therefore, each attended pair is one that the physics predicts to be well correlated. By contrast, applying the same keep set to a JST encoder spreads its quadratic attention budget over all pairs of retained tokens, including pairs that differ on both axes simultaneously. For such cross-axis pairs, the autocorrelation factorizes as $R_{\mathrm{t}}(\Delta t)\,R_{\mathrm{sf}}(\Delta f,\Delta s)$, which can decay rapidly through either factor and so carries comparatively little signal. FST therefore extracts more information per unit of  token budget than JST on the same visible set, allowing aggressive masking to remain useful rather than degrading into mixing across weakly correlated lags.

\subsection{Patch-normalized reconstruction with an auxiliary scale loss}
The power of wireless channel spans an enormous dynamic range across samples. Path loss and shadowing typically vary by tens of dB between channels in the same training set, while small-scale multipath fading, the relevant structure for beam selection and many other downstream tasks, occupies a much finer amplitude scale. A reconstruction loss computed in raw amplitude would therefore be dominated by high-power patches, and the network would minimize it by fitting large-scale trends while leaving the multipath geometry largely unlearned. We address this by normalizing each patch by its own mean and variance before computing the reconstruction loss. Writing $\mathbf{p}_{b,i} \in \mathbb{R}^{D_\mathrm{p}}$ for the $i$-th patch of sample $b$ and $\mu_{b,i}$, $\sigma^2_{b,i}$ for its empirical mean and variance, respectively, the reconstruction loss over the masked set $\mathcal{M}$ is
\begin{equation}
  \mathcal{L}_{\mathrm{recon}} = \frac{1}{|\mathcal{M}|} \sum_{(b,i) \in \mathcal{M}} \left\| \hat{\mathbf{p}}_{b,i} - \frac{\mathbf{p}_{b,i} - \mu_{b,i}\mathbf{1}}{\sqrt{\sigma^2_{b,i} + \epsilon_\mathrm{r}}} \right\|_2^2,
  \label{eq:recon-loss}
\end{equation}
where $\epsilon_\mathrm{r}$ is a small stability constant. By cancelling per-patch amplitude, this loss forces the encoder to represent the inter-patch fading structure and correlations rather than the sample-level power that dominates the raw signal. This also removes a trivial shortcut in which the network minimizes the loss by predicting patch means in raw space. As a result, dividing out per-patch amplitude discards the large-scale fading signature (path loss and shadowing) that certain downstream tasks rely on. This motivates us to introduce the auxiliary scale loss as follows.

For each raw patch, we form the target
\begin{equation}
  \mathbf{s}_{b,i} = \begin{pmatrix} \mu_{b,i} \\ \log(\sigma^2_{b,i} + \epsilon_{\mathrm{s}}) \end{pmatrix},
  \label{eq:scale-target}
\end{equation}
representing the variance in the log scale for numerical stability. We ask the model to predict $\mathbf{s}_{b,i}$ from both encoder and decoder token features. The encoder-side predictions $\hat{\mathbf{s}}^{\mathrm{enc}}_{b,i}$ are self-supervised on the visible set $\mathcal{K}$. Therefore, the encoder itself learns to carry large-scale statistics in its latent. The predictions at the decoder-side, $\hat{\mathbf{s}}^{\mathrm{dec}}_{b,i}$, are self-supervised on the masked set $\mathcal{M}$. As a result, the reconstruction path also recovers the scale. Although the decoder-side term is supervised at the decoder, its gradient propagates back through the decoder into the encoder. Therefore, it additionally shapes encoder visible-token representations to encode scale information in a form the decoder can use to predict scales at masked positions. The two corresponding loss terms are
\begin{equation}
  \begin{aligned}
    \mathcal{L}_{\mathrm{scale,enc}} &= \frac{1}{|\mathcal{K}|} \sum_{(b,i) \in \mathcal{K}} \left\| \hat{\mathbf{s}}^{\mathrm{enc}}_{b,i} - \mathbf{s}_{b,i} \right\|_2^2, \\
    \mathcal{L}_{\mathrm{scale,dec}} &= \frac{1}{|\mathcal{M}|} \sum_{(b,i) \in \mathcal{M}} \left\| \hat{\mathbf{s}}^{\mathrm{dec}}_{b,i} - \mathbf{s}_{b,i} \right\|_2^2,
  \end{aligned}
  \label{eq:scale-losses}
\end{equation}
and the full pretraining objective is
\begin{equation}
  \mathcal{L}_{\mathrm{pretrain}} = \mathcal{L}_{\mathrm{recon}} + \lambda_{\mathrm{enc}} \mathcal{L}_{\mathrm{scale,enc}} + \lambda_{\mathrm{dec}} \mathcal{L}_{\mathrm{scale,dec}}.
  \label{eq:pretrain-loss}
\end{equation}

\begin{figure}[b]
  \centering
  \includegraphics[width=\linewidth]{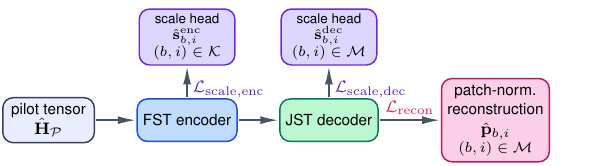}
  \caption{PilotWiMAE pretraining flow and loss groupings. Auxiliary scale heads attach to encoder and decoder features. Reconstruction and decoder-side scale use masked patches $\mathcal{M}$, while encoder-side scale uses visible patches $\mathcal{K}$.}
  \label{fig:pretrain-losses}
\end{figure}

Fig.~\ref{fig:pretrain-losses} summarizes the pretraining wiring. The pilot tensor feeds the FST encoder and JST decoder. Patch-normalized reconstruction predicts $\hat{\mathbf{p}}_{b,i}$ on $(b,i)\in\mathcal{M}$, while scale heads predict $\hat{\mathbf{s}}^{\mathrm{enc}}_{b,i}$ on $\mathcal{K}$ and $\hat{\mathbf{s}}^{\mathrm{dec}}_{b,i}$ on $\mathcal{M}$, matching the supervision sets in \eqref{eq:recon-loss}, \eqref{eq:scale-losses}, and \eqref{eq:pretrain-loss}.

\subsection{Noise-robust pretraining curriculum}
During pretraining, the sparse masked input is corrupted with AWGN while the reconstruction and scale targets are computed against the clean channel. This prepares the encoder for deployment, where it processes sparse pilot-pattern inputs under the corresponding noise regime.

More precisely, let $e \in \{0, \ldots, E-1\}$ index the pretraining epoch, let $s_0$ denote the initial lower bound for the SNR range in dB, and let $s_{\max}$ denote the upper bound in dB. The lower bound follows a cosine schedule that anneals from $s_0$ down to $0$ dB,
\begin{equation}
  s_{\min}(e) = \frac{s_0}{2}\left(1 + \cos\left(\frac{\pi e}{E-1}\right)\right),
  \label{eq:snr-curriculum}
\end{equation}
so that early epochs concentrate on higher-SNR observations and later epochs progressively expose the model to lower SNRs, widening the pretraining distribution as the encoder stabilizes. Then, we sample
\begin{equation}
  \mathrm{SNR}_{b,\mathrm{dB}} \sim \mathcal{U}[s_{\min}(e), s_{\max}].
  \label{eq:snr-sample}
\end{equation}
For each sample $b$, we measure the channel power $P_b$ on the kept patches that the encoder actually consumes. Then, the linear SNR sets the noise variance $\sigma^2_b = P_b / \mathrm{SNR}_{b,\mathrm{lin}}$. Circularly symmetric complex Gaussian noise of variance $\sigma^2_b$ is drawn independently and added to the visible elements only, while the reconstruction and scale losses are evaluated against statistics computed from the clean full-grid channel $\mathbf{H}_b$.

\subsection{Two-phase pretraining schedule}
\label{sec:approach:decoder-phase}
Masked reconstruction couples encoder representations to the paired decoder. When the decoder is small, gradients emphasize encoder-side integration of context from visible pilots. When it is large, the decoder can partly compensate for weaker latents, weakening pressure on the encoder to encode useful geometry.

\paragraph{Phase~1 (joint encoder and decoder) pretraining}
We pretrain with a deliberately shallow JST decoder of $L_{\mathrm{JST}}$ layers and the full objective~\eqref{eq:pretrain-loss}. This makes encoder latents well-suited for downstream pilot-native tasks while the paired decoder stays intentionally capacity-limited.

\paragraph{Phase~2 (decoder-centric) pretraining}
We discard the shallow decoder weights, freeze the encoder, attach a freshly initialized decoder with greater reconstruction capacity (for example more layers, feed-forward width, or attention heads), and pretrain only this decoder on reconstruction only. As a result, encoder representations stay fixed while the new decoder absorbs the residual mapping from frozen tokens to dense channel patches. Procedure-wise, Phase~2 resembles downstream adaptation (fixed backbone, new head), but the objective remains masked self-supervised reconstruction under structured random masking rather than a supervised objective on a fixed pilot grid. The resulting reconstruction will be agnostic to the pilot pattern. This is fundamentally different 
from conventional channel estimation, where the decoder is supervised to learn a mapping from a fixed pilot pattern to full CSI.

The two-phase schedule produces a single pretrained encoder-decoder pair. Downstream tasks that do not require dense CSI consume only the encoder's features. On the other hand, applications that require accurate dense CSI additionally use the decoder output. Pretraining parameters appear in Section~\ref{sec:experiments:pretraining} and Table~\ref{tab:pretrain-hparams}.

\section{Experiments}
\label{sec:experiments}

We pretrain PilotWiMAE on a ray-tracing channel dataset at $3.5$\,GHz. Then, we evaluate transfer without task-specific fine-tuning on held-out test data. For cross-frequency beam selection and channel characterization, our evaluation includes in-distribution (ID) and out-of-distribution (OOD) settings. We cover frequency mismatch alone (ID, $3.5$ to $28$\,GHz on pretraining cities) and combined frequency-plus-city mismatch (OOD, $3.5$ to $28$\,GHz on the held-out city). For channel estimation, we report dense-channel recovery at $3.5$\,GHz on the held-out city, isolating scene mismatch under matched carrier.

\subsection{Dataset}
\label{sec:experiments:dataset}

We create a ray-tracing channel dataset using our in-house generation pipeline CSIGen, built on Sionna \cite{Sionna2025}, across urban deployment scenarios. Each scenario uses six base stations and includes both LoS and NLoS links, with channel tensors generated at $3.5$\,GHz and $28$\,GHz under a shared simulation protocol. The pretraining split uses Boston, New York City, San Francisco, and Chicago. For evaluation, we use unseen channels from these same cities for ID testing. We also use Los Angeles as a held-out city for OOD testing. Combined with the $3.5$ to $28$\,GHz carrier shift, this setup isolates frequency-only transfer (ID) from frequency-plus-scene transfer (OOD). Per-city channel counts and scene sizes are summarized in Appendix~\ref{app:dataset}. Table~\ref{tab:dataset-sim-params} reports the Sionna-based generation parameters used across cities.

\begin{table}[b]
  \centering
  \caption{Sionna-based dataset generation parameters}
  \label{tab:dataset-sim-params}
  \footnotesize
  \begin{tabular}{ll}
    \toprule
    Parameter & Value \\
    \midrule
    Carrier frequencies & $3.5$\,GHz and $28$\,GHz \\
    Number of BSs & 6 (per scenario) \\
    BS sectors / mechanical downtilt & 6 sectors / $10^\circ$ \\
    BS height above rooftop & $10$\,m vertical offset \\
    BS transmit power & 43\,dBm per array \\
    TX array & $4\times 8$, vertical pol., $0.5\lambda$ spacing \\
    TX element pattern & TR~38.901 \cite{3GPP38901} \\
    RX array & $1\times 1$, isotropic, vertical pol. \\
    UE height offset & 1.5\,m above ground \\
    UE mobility & $8$-$30$\,m/s \\
    OFDM grid & 14 symbols, 32 subcarriers \\
    Subcarrier spacing & 30\,kHz \\
    \bottomrule
  \end{tabular}
\end{table}

For pilot-only inference and evaluation, we use a fixed pilot placement on the OFDM grid with time slot indices $\mathcal{T}_{\mathrm{p}} = \{2,11\}$ and subcarrier indices $\mathcal{F}_{\mathrm{p}} = \{0,1,2,3,8,9,10,11,16,17,18,19,24,25,26,27\}$, observing all $N_{\mathrm{h}}N_{\mathrm{v}}$ antennas at each pilot resource element. The time-domain placement follows 5G NR PDSCH DMRS Mapping Type A~\cite{3GPP_TS38.211}, with Symbol 2 as the front-loaded DMRS and Symbol 11 as the additional DMRS, while the frequency-domain placement uniformly tiles the band. This yields $|\mathcal{T}_{\mathrm{p}}| \times |\mathcal{F}_{\mathrm{p}}| = 32$ pilots among $|\mathcal{T}| \times |\mathcal{F}| = 448$ resource elements. The same mask is used across all downstream evaluations for a controlled comparison, with no pattern-specific fine-tuning. Fig.~\ref{fig:pilot-mask} in Appendix~\ref{app:pilot-pattern} illustrates the pattern on the time-frequency grid.

\subsection{Pretraining}
\label{sec:experiments:pretraining}

We pretrain PilotWiMAE at $3.5$\,GHz under the two-phase schedule of Section~\ref{sec:approach:decoder-phase}, with $500$ phase-1 epochs followed by $200$ phase-2 epochs. We instantiate the FST encoder with $L_{\mathrm{FST}}=3$ blocks and eight heads, the JST decoder with $L_{\mathrm{JST}}=2$ layers and four heads in Phase~1, eight heads in Phase~2, and model dimension $d=128$ throughout. Phase~1 applies aggressive factorized masking ($n_{\mathrm{k}}=2$, $\rho_{\mathrm{k}}=0.1$, $\sim 99\%$ overall), while Phase~2 uses a milder budget ($n_{\mathrm{k}}=4$, $\rho_{\mathrm{k}}=0.75$, $\sim 79\%$ overall). The asymmetry follows from the different objectives of the two phases. Phase~1 shapes the encoder representation, where forcing prediction from very few visible tokens compels integration of long-range structure across the grid. In our experiments, this improves downstream accuracy as the mask ratio is pushed toward the aggressive regime. Instead, Phase~2 shapes the decoder with the encoder frozen, where loosening the budget to $\sim 79\%$ supplies enough visible context for accurate dense recovery while still randomizing the kept set to preserve pilot-pattern agnosticity. Table~\ref{tab:pretrain-hparams} lists the remaining architecture and optimization details.

\begin{table}[b]
  \centering
  \caption{Pretraining architecture and optimization}
  \label{tab:pretrain-hparams}
  \footnotesize
  \begin{tabularx}{\linewidth}{@{}>{\raggedright\arraybackslash}p{0.34\linewidth} >{\raggedright\arraybackslash}X@{}}
    \toprule
    Setting & Value \\
    \midrule
    Factorized masking & Phase~1: $n_{\mathrm{k}}=2$, $\rho_{\mathrm{k}}=0.1$ ($\sim 99\%$ overall) Phase~2: $n_{\mathrm{k}}=4$, $\rho_{\mathrm{k}}=0.75$ ($\sim 79\%$ overall)\\
    Model dimension & $d=128$ \\
    FFN expansion & $\kappa_{\mathrm{ff}}=4$ \\
    Pos.\ enc.\ scale & $\alpha_{\mathrm{pe}} = 0.01$ at start \\
    Encoder depth / heads & $L_{\mathrm{FST}}=3$ blocks / 8 heads \\
    Decoder type & JST decoder \\
    Decoder depth / heads & $L_{\mathrm{JST}}=2$ layers / 4 heads (phase~1), 8 heads (phase~2) \\
    Optimizer & AdamW, $\beta=(0.9,0.999)$, wd $0.005$ \\
    Epochs & $500$ (phase~1) + $200$ (phase~2) \\
    LR schedule & Cosine decay in each phase \\
    Phase-1 LR & $\eta_{\mathrm{start}}=5\!\times\!10^{-4}$, $\eta_{\min}=5\!\times\!10^{-6}$ \\
    Phase-2 LR & $\eta_{\mathrm{start}}=10^{-4}$, $\eta_{\min}=10^{-6}$ \\
    Aux. scale losses & $\lambda_{\mathrm{enc}}=\lambda_{\mathrm{dec}}=0.05$ (phase~1 only) \\
    SNR curriculum & \eqref{eq:snr-curriculum}-\eqref{eq:snr-sample}, phase~1 only ($E{=}500$) \\
    Noise setup & $s_0=40$\,dB, $s_{\max}=40$\,dB \\
    Batch size & 512 \\
    Precision / clipping & Mixed precision, gradient clip at 1.0 \\
    \bottomrule
  \end{tabularx}
\end{table}

\subsection{Tasks}
\label{sec:experiments:tasks}

We evaluate three downstream tasks. Cross-frequency beam selection and channel characterization (LoS/NLoS classification) are evaluated under the cross-band transfer protocol ($3.5$ to $28$\,GHz). Beam selection tests whether the learned representation preserves directional structure across bands, while channel characterization tests whether it preserves propagation-state semantics under frequency-dependent channel statistics. For both, we use frozen pretrained representations and k-Nearest Neighbors (kNN) evaluation without task-specific fine-tuning. For beam selection, labels are defined using a DFT codebook \cite{3GPP38214}. Channel estimation is evaluated at matched carrier ($3.5$\,GHz) on the held-out city and reads out dense-channel reconstructions from the decoder.

\subsubsection{Shared transfer protocol}

Beam selection and channel characterization use the same frozen-feature transfer protocol:
\begin{itemize}
  \item The representation encoder is pretrained at $3.5$\,GHz and then frozen.
  \item Evaluation is performed at $28$\,GHz without task-specific fine-tuning.
  \item Features are evaluated with a common kNN protocol for all compared methods (10 disjoint folds), using mean-pooled encoder representations.
  \item In each fold, kNN is fit on $90\%$ of samples and evaluated on the $10\%$ held-out.
  \item We report mean and standard deviation over cross-validation folds.
\end{itemize}
We use kNN as the readout because it is non-parametric and adds no trainable parameters on top of the frozen encoder. kNN's accuracy reflects how well the learned feature geometry already groups channels by task-relevant similarity. Even a linear probe would improve the results while adding a fresh inductive bias of its own (a learned linear separator over the feature space). This can compensate for representations whose neighborhood structure is itself uninformative. By contrast, kNN succeeds only when nearby points in the learned representation also share the downstream label, which is a property that a good self-supervised representation is supposed to have. Therefore, kNN is a better choice for evaluating the performance. 

\subsubsection{kNN details}
We use $k=20$ with cosine-distance-based weighted voting \cite{Dudani1976}. Mean-pooled encoder features are passed to kNN without external feature normalization. For cosine distance, L2 normalization is applied. For all models and both full-channel and pilot-only inputs, mean pooling over encoded patches yields a $d=128$ representation.

\subsubsection{Cross-frequency beam selection}
\label{sec:experiments:beam-selection}

Beam selection evaluates whether pretrained representations preserve the directional structure across frequency bands, with performance reported as top-3 beam-selection accuracy.
For a uniform planar array (UPA) with $N_{\mathrm{h}}$ columns and $N_{\mathrm{v}}$ rows and half-wavelength spacing in both dimensions, let $K_{\mathrm{h}}$ and $K_{\mathrm{v}}$ denote the number of angular bins along horizontal and vertical axes. Along each axis, we use either an \emph{oversampled} DFT grid (which subsumes the critically sampled grid as the special case $K_{\mathrm{h}}=N_{\mathrm{h}}$, $K_{\mathrm{v}}=N_{\mathrm{v}}$, yielding $M=N_{\mathrm{h}}N_{\mathrm{v}}$ codewords) or an \emph{undersampled} grid:
\begin{equation}
  \begin{aligned}
    K_{\mathrm{h}} &=
    \begin{cases}
      O_{\mathrm{h}}N_{\mathrm{h}}, & \text{oversampled } (O_{\mathrm{h}}\geq 1),\\[4pt]
      N_{\mathrm{h}}/U_{\mathrm{h}}, & \text{undersampled } (U_{\mathrm{h}}\geq 1,\; U_{\mathrm{h}}\mid N_{\mathrm{h}}),
    \end{cases}\\[6pt]
    K_{\mathrm{v}} &=
    \begin{cases}
      O_{\mathrm{v}}N_{\mathrm{v}}, & \text{oversampled } (O_{\mathrm{v}}\geq 1),\\[4pt]
      N_{\mathrm{v}}/U_{\mathrm{v}}, & \text{undersampled } (U_{\mathrm{v}}\geq 1,\; U_{\mathrm{v}}\mid N_{\mathrm{v}}),
    \end{cases}
  \end{aligned}
  \label{eq:codebook-K}
\end{equation}
where $a \mid b$ denotes that $a$ divides $b$, $O_{\mathrm{h}},O_{\mathrm{v}},U_{\mathrm{h}},U_{\mathrm{v}}$ are positive integers that define over- and undersampling factors. Only one branch of~\eqref{eq:codebook-K} is active per axis. Undersampled codebooks remain uniform in $[0,2\pi)$ on each ring. The total codebook size is $M=K_{\mathrm{h}}K_{\mathrm{v}}$, and codewords are Kronecker products of 1D steering vectors,
\begin{equation}
  \begin{aligned}
    \relax[\mathbf{a}_{\mathrm{h}}(m_{\mathrm{h}})]_{n} &= \frac{1}{\sqrt{N_{\mathrm{h}}}}\exp\!\bigl(\mathrm{j}\, n\, \phi_{m_{\mathrm{h}}}\bigr),\\
    &\quad \phi_{m_{\mathrm{h}}}=\frac{2\pi\,m_{\mathrm{h}}}{K_{\mathrm{h}}},\quad n=0,\ldots,N_{\mathrm{h}}-1,\\[-2pt]
    \relax[\mathbf{a}_{\mathrm{v}}(m_{\mathrm{v}})]_{n} &= \frac{1}{\sqrt{N_{\mathrm{v}}}}\exp\!\bigl(\mathrm{j}\, n\, \psi_{m_{\mathrm{v}}}\bigr),\\
    &\quad \psi_{m_{\mathrm{v}}}=\frac{2\pi\,m_{\mathrm{v}}}{K_{\mathrm{v}}},\quad n=0,\ldots,N_{\mathrm{v}}-1,
  \end{aligned}
  \label{eq:codebook-steer}
\end{equation}
\begin{equation}
  \mathbf{w}_{m_{\mathrm{h}},m_{\mathrm{v}}} = \mathbf{a}_{\mathrm{v}}(m_{\mathrm{v}})\otimes \mathbf{a}_{\mathrm{h}}(m_{\mathrm{h}}),
  \label{eq:codebook-kron}
\end{equation}
with $m_{\mathrm{h}}\in\{0,\ldots,K_{\mathrm{h}}-1\}$, $m_{\mathrm{v}}\in\{0,\ldots,K_{\mathrm{v}}-1\}$, flat index $m=m_{\mathrm{v}}K_{\mathrm{h}}+m_{\mathrm{h}}$, $\mathbf{w}_m\equiv \mathbf{w}_{m_{\mathrm{h}},m_{\mathrm{v}}}$, and $\otimes$ denotes the Kronecker product. The $1/\sqrt{N_{\mathrm{h}}}$ and $1/\sqrt{N_{\mathrm{v}}}$ scaling in~\eqref{eq:codebook-steer} keeps $\|\mathbf{w}_m\|_2=1$ for any $(K_{\mathrm{h}},K_{\mathrm{v}})$. Given channel vectors $\mathbf{h}_{t,f}$, a single beam label per frame is assigned by maximizing average beam gain over subcarriers and slots:
\begin{equation}
  m^\star = \arg\max_{m\in\{0,\ldots,M-1\}} \frac{1}{T\,N_{\mathrm{sc}}}\sum_{t=0}^{T-1}\sum_{f=0}^{N_{\mathrm{sc}}-1}\left|\mathbf{w}_m^{\mathrm{H}}\mathbf{h}_{t,f}\right|^2,
  \label{eq:beam-label}
\end{equation}
since the angular structure is approximately constant within an OFDM frame. Unless stated otherwise, we use $(N_{\mathrm{h}},N_{\mathrm{v}})=(8,4)$. The main beam-selection plots fix $M{=}128$, obtained from~\eqref{eq:codebook-K} via $(O_{\mathrm{h}},O_{\mathrm{v}})=(2,2)$ on the oversampled branch. Fig.~\ref{fig:beam-ood-cdb-scaling} additionally sweeps other $M$ cardinalities, including the critically sampled construction $(O_{\mathrm{h}},O_{\mathrm{v}})=(1,1)$, undersampled $(U_{\mathrm{h}},U_{\mathrm{v}})$ constructions at small $M$, and larger oversampled $(O_{\mathrm{h}},O_{\mathrm{v}})$ at high $M$. The legend entries list the active factors per curve.

\subsubsection{Channel characterization}

Channel characterization evaluates whether pretrained representations preserve propagation-state semantics under the same $3.5$ to $28$\,GHz transfer. The binary LoS/NLoS label is the ray-tracer's geometric LoS indicator (whether the direct base-station-to-UE path is unobstructed) and is therefore carrier-independent. A given link carries the same label at $3.5$\,GHz and $28$\,GHz, while its small-scale and large-scale statistics differ across carriers. Performance is reported by LoS classification accuracy in both ID and OOD settings. The dataset details for different test cities are provided in Appendix~\ref{app:dataset}.

\subsubsection{Compared methods}

\paragraph{Beam selection and channel characterization}
For these two tasks, we compare the following methods:
\begin{itemize}
  \item \emph{Supervised baseline}: an FST encoder followed by a linear classification head, trained end-to-end on full-channel inputs under cross-entropy loss. For kNN evaluation, the classifier is discarded and mean-pooled encoder features are used.
  \item \emph{Self-supervised JST baseline}: a JST encoder paired with a JST decoder of matching capacity, pretrained with patch-normalized reconstruction alone (no AWGN curriculum, no auxiliary scale heads).
  \item \emph{PilotWiMAE ablations}: three reduced variants that share PilotWiMAE's FST encoder and phase-1 JST decoder but toggle PilotWiMAE's two pretraining ingredients on and off. \emph{FST} uses patch-normalized reconstruction alone. \emph{FST+scale} adds the auxiliary encoder- and decoder-side scale heads. \emph{FST+noise} adds the AWGN SNR curriculum.
  \item \emph{PilotWiMAE} (also denoted \emph{FST+noise+scale} in the figures): the full proposed method, combining the FST encoder, JST decoder, AWGN curriculum, and auxiliary scale heads.
\end{itemize}
All compared methods are pretrained, or trained (for the supervised baseline), on the same $3.5$\,GHz pretraining split as PilotWiMAE (Section~\ref{sec:experiments:dataset}). To isolate the effect of the factorized inductive bias, the JST and FST encoders are matched in capacity. They share $d=128$, eight attention heads, and the same number of attention sublayers ($L_{\mathrm{FST}}=3$ FST blocks contribute six attention sublayers, equal to the six layers of the JST encoder), and consequently have the same number of trainable parameters. The self-supervised JST baseline pairs its encoder with a JST decoder of the same depth and heads as the PilotWiMAE phase-1 decoder. Therefore, any gap between the JST baseline and the FST ablation reflects the attention factorization in the encoder rather than a difference in the choice of the decoder. The supervised baseline shares the same FST encoder configuration as PilotWiMAE and replaces the JST decoder with a task-specific linear head. Detailed configurations of the JST self-supervised baseline and the supervised baseline are provided in Appendix~\ref{app:baseline-configs}.

\paragraph{Channel estimation}
For dense-channel recovery, PilotWiMAE is compared against classical estimators and supervised encoder-decoder baselines:
\begin{itemize}
  \item Linear interpolation: a lightweight classical baseline.
  \item \emph{LMMSE practical}: a Kronecker-structured LMMSE estimator whose second-order channel statistics are estimated from the $3.5$\,GHz pretraining split and evaluated on Los Angeles test channels at $3.5$\,GHz, isolating scene mismatch under matched carrier.
  \item \emph{LMMSE gold}: the same Kronecker structure with statistics computed on Los Angeles itself for both covariance estimation and evaluation, representing an oracle matched to the deployment scenario.
  \item \emph{Supervised FST}: the FST encoder configuration matched to PilotWiMAE paired with the same JST decoder used in PilotWiMAE's Phase~2, supervised end-to-end against the full channel tensor with the fixed pilot pattern of Appendix~\ref{app:pilot-pattern} at the input.
  \item \emph{Supervised JST}: the same end-to-end setup as supervised FST but with a standard JST encoder, testing whether the factorized restriction remains expressive enough for the regression task of channel estimation.
\end{itemize}
We adopt the Kronecker covariance model for both LMMSE references because it parallels the factorized inductive bias of the FST encoder and, equally important, makes the LMMSE computation feasible at our grid sizes.

\subsection{Results}

\begin{figure}
  \centering
  \includegraphics[width=\linewidth]{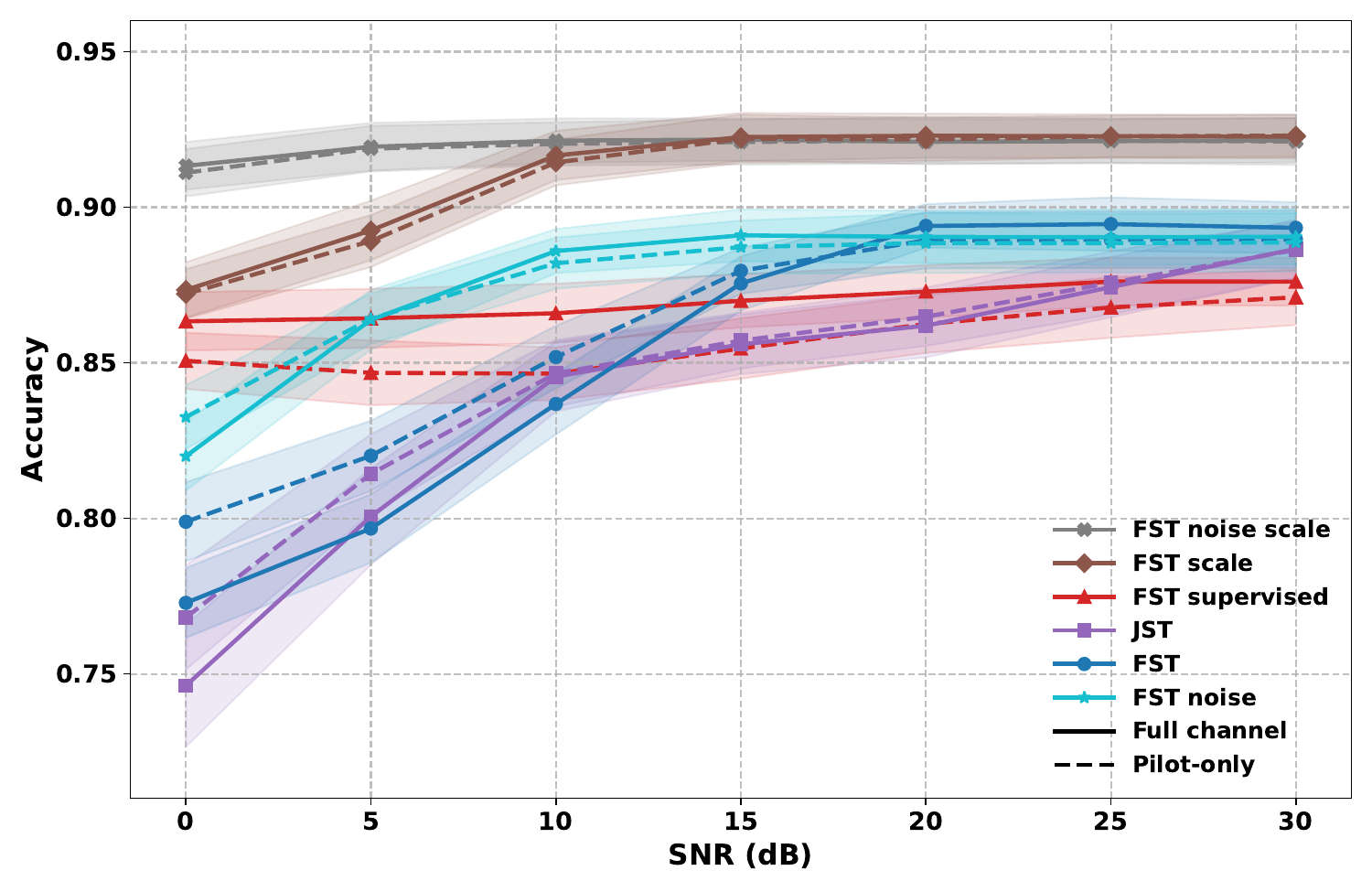}
  \caption{Channel characterization (LoS accuracy) in-distribution at 28\,GHz.}
  \label{fig:los-id-28}
\end{figure}

\begin{figure}
  \centering
  \includegraphics[width=\linewidth]{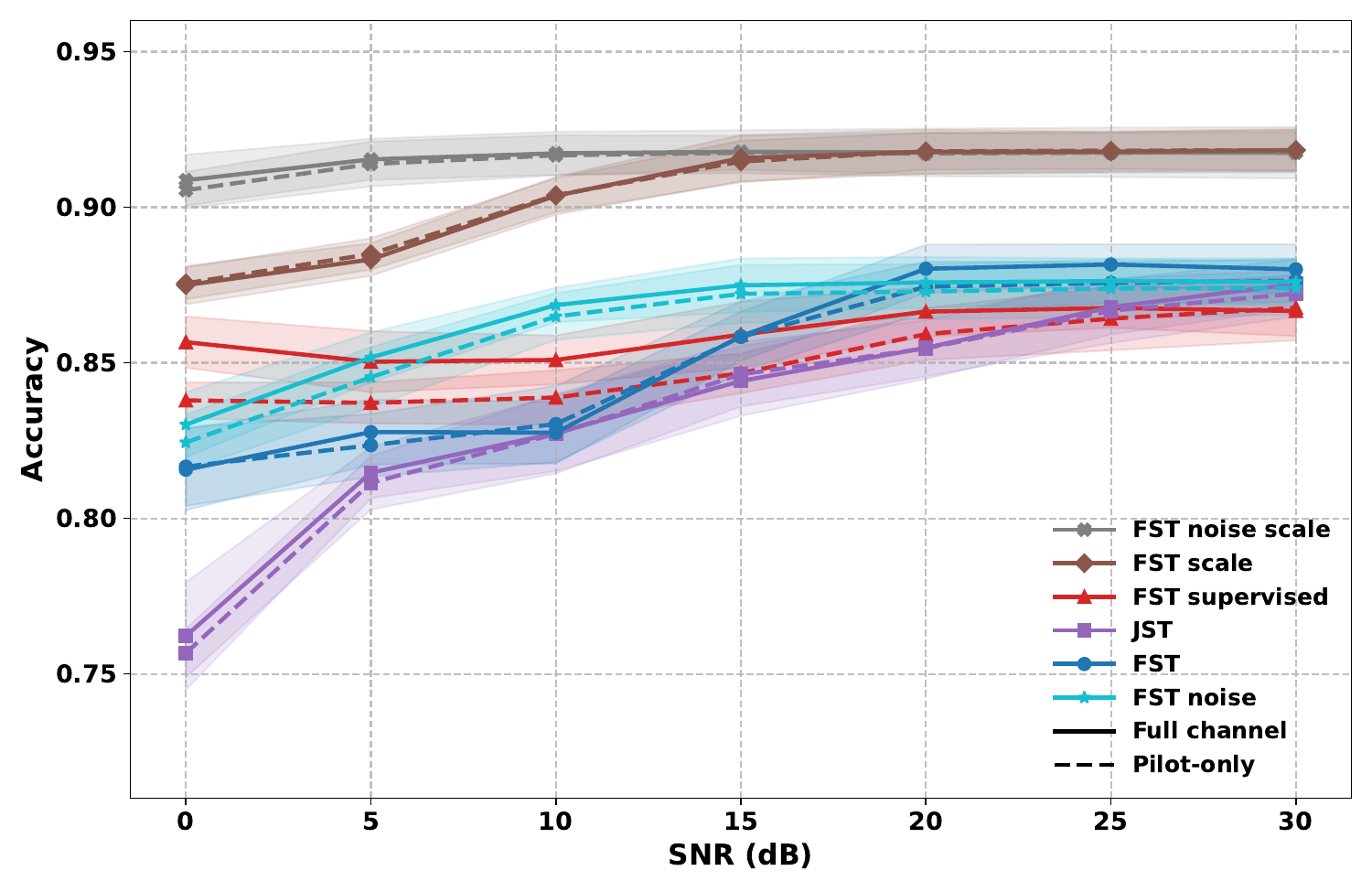}
  \caption{Channel characterization (LoS accuracy) out-of-distribution at 28\,GHz.}
  \label{fig:los-ood-28}
\end{figure}

Across beam selection and channel characterization, the factorized encoder family is consistently more robust than JST under both ID and OOD transfers, especially in lower-SNR regimes. Noise-robust pretraining improves stability across SNR, and the auxiliary scale objective is most beneficial for channel characterization, where large-scale fading cues are directly informative. Its effect on beam selection is smaller. Overall, PilotWiMAE, depicted as the FST noise scale, provides the best performance in Figs.~\ref{fig:beam-id-28}-\ref{fig:los-ood-28}, while pilot-only inference remains competitive with full-channel inputs despite using a substantially smaller observation space. In channel estimation, the normalized mean squared error (NMSE) of dense-channel recovery is improved by increasing the decoder depth. PilotWiMAE achieves competitive performance against classical and supervised pilot-pattern-specific encoder-decoder baselines despite being pretrained without commitment to a specific pilot pattern (Figs.~\ref{fig:ce-ood-decoder-depth}-\ref{fig:ce-ood-comparative}).
\begin{figure}[t]
  \centering
  \includegraphics[width=\linewidth]{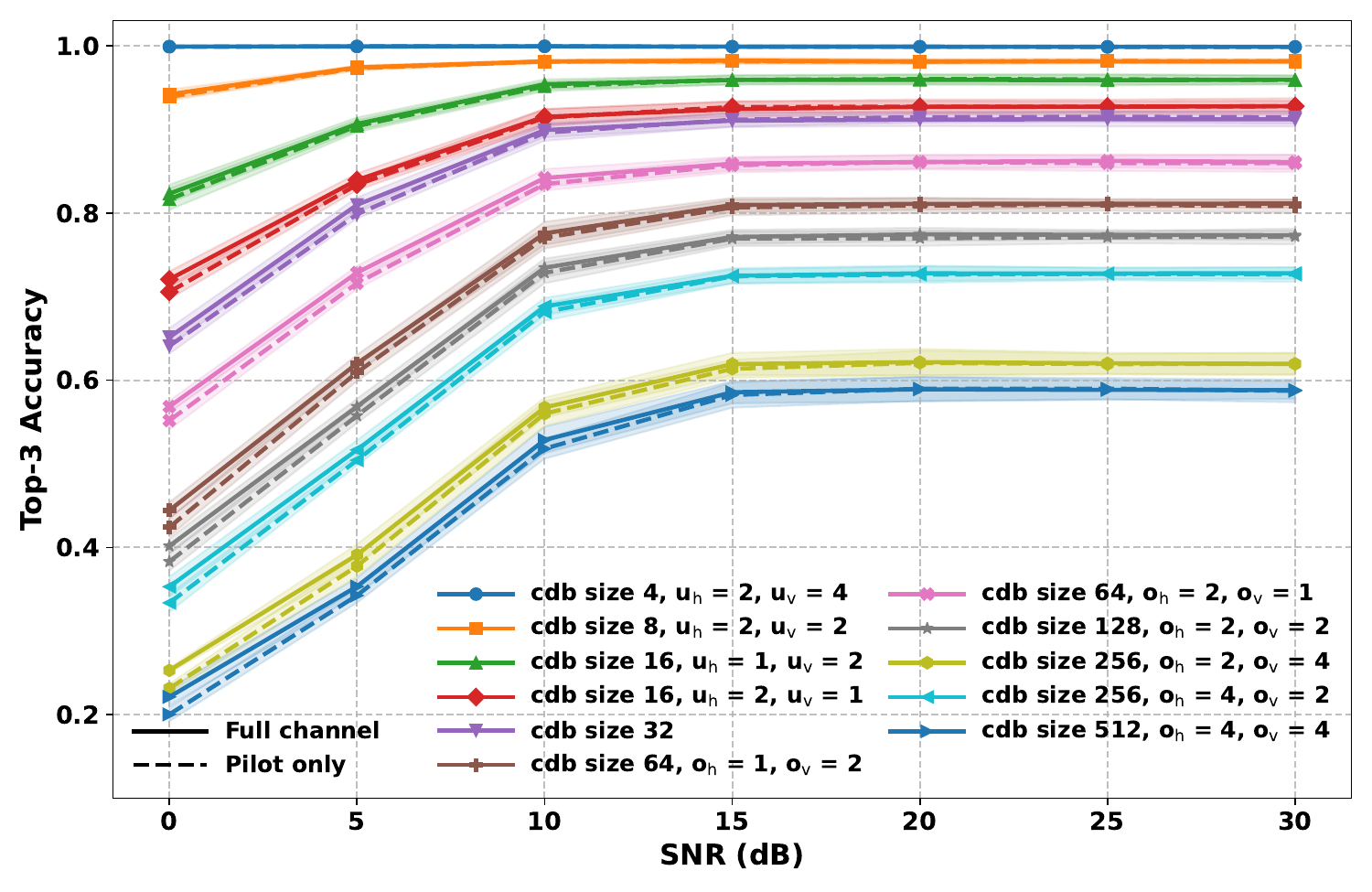}
  \caption{OOD (28\,GHz, Los Angeles): top-3 beam-selection accuracy vs SNR for PilotWiMAE (FST+noise+scale) using DFT codebooks and horizontal-vertical Kronecker factorizations that match the codebook size $M$.}
  \label{fig:beam-ood-cdb-scaling}
\end{figure}
\subsubsection{Cross-frequency beam selection}

In the Los Angeles OOD split at $28$\,GHz, Fig.~\ref{fig:beam-ood-cdb-scaling} reports the top-3 accuracy versus SNR in a wide range of codebook sizes, including multiple horizontal-vertical Kronecker constructions that yield the same codebook size $M$. For each construction, we evaluate the same frozen encoder features from PilotWiMAE under pilot-only and full-channel inputs with an otherwise identical kNN readout. The pilot-versus-full gap depends strongly on the SNR, typically widest at low SNR and tightening toward high SNR, whereas across codebook sizes and tilings we do not observe a simple monotone trend with $M$ alone. In particular, at matched $M$, the horizontal-vertical factorization itself shifts accuracy by several points, with constructions that allocate more bins to the horizontal axis (e.g., $u_h{<}u_v$ on the undersampled branch or $o_h{>}o_v$ on the oversampled branch) consistently outperforming their transposed counterparts. This suggests that, given our scene setup, the representation captures horizontal directionality more sharply than vertical. As a result, undersampling the vertical axis or oversampling the horizontal one preserves more of the discriminative angular structure at a fixed budget. Head-to-head comparisons to supervised and self-supervised baselines at a representative fine codebook ($M{=}128$) are discussed next.

\begin{figure}
  \centering
  \includegraphics[width=\linewidth]{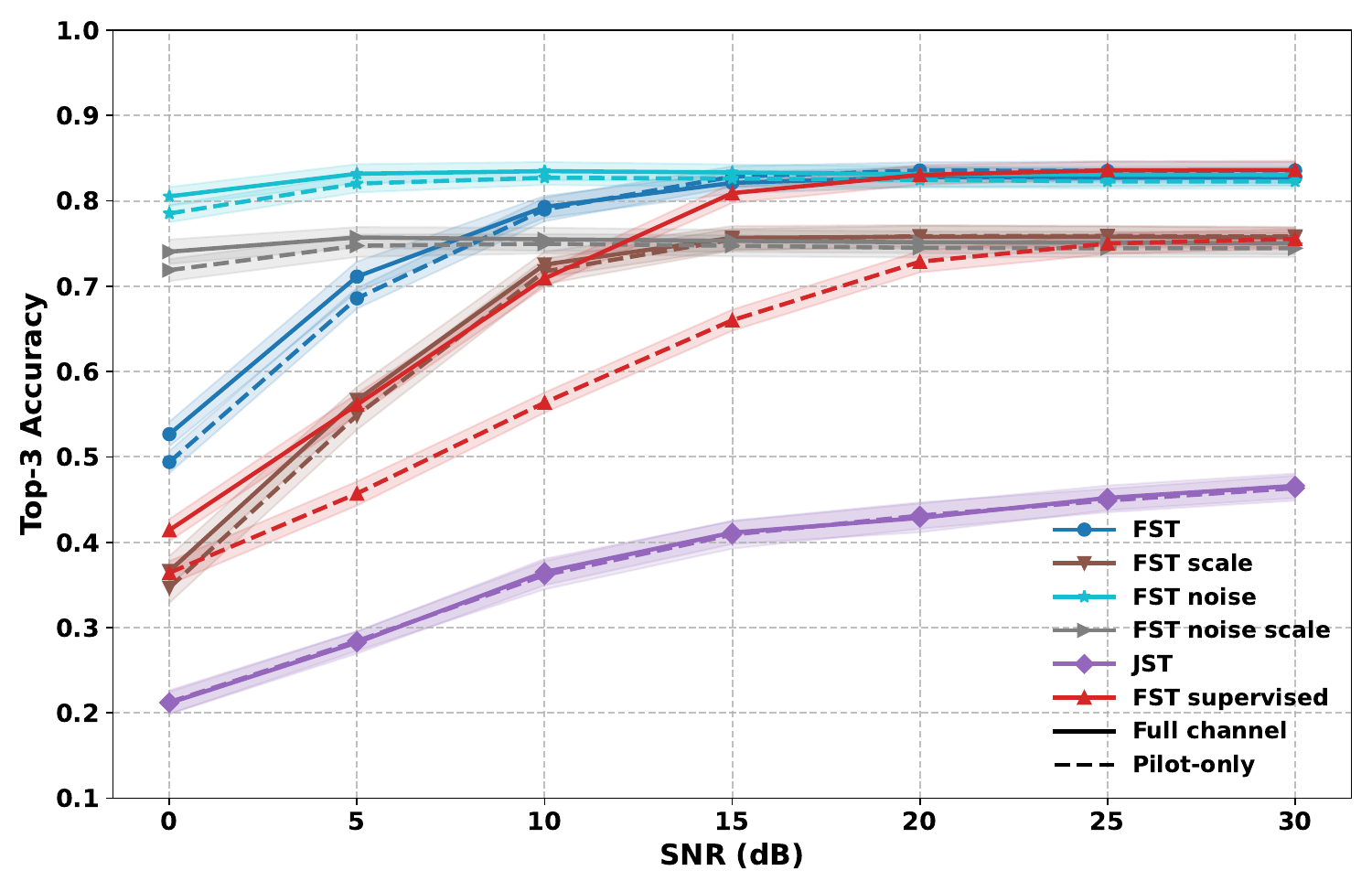}
  \caption{Cross-frequency beam selection (top-3 accuracy) in-distribution at 28\,GHz with codebook size $M=128$.}
  \label{fig:beam-id-28}
\end{figure}

\begin{figure}
  \centering
  \includegraphics[width=\linewidth]{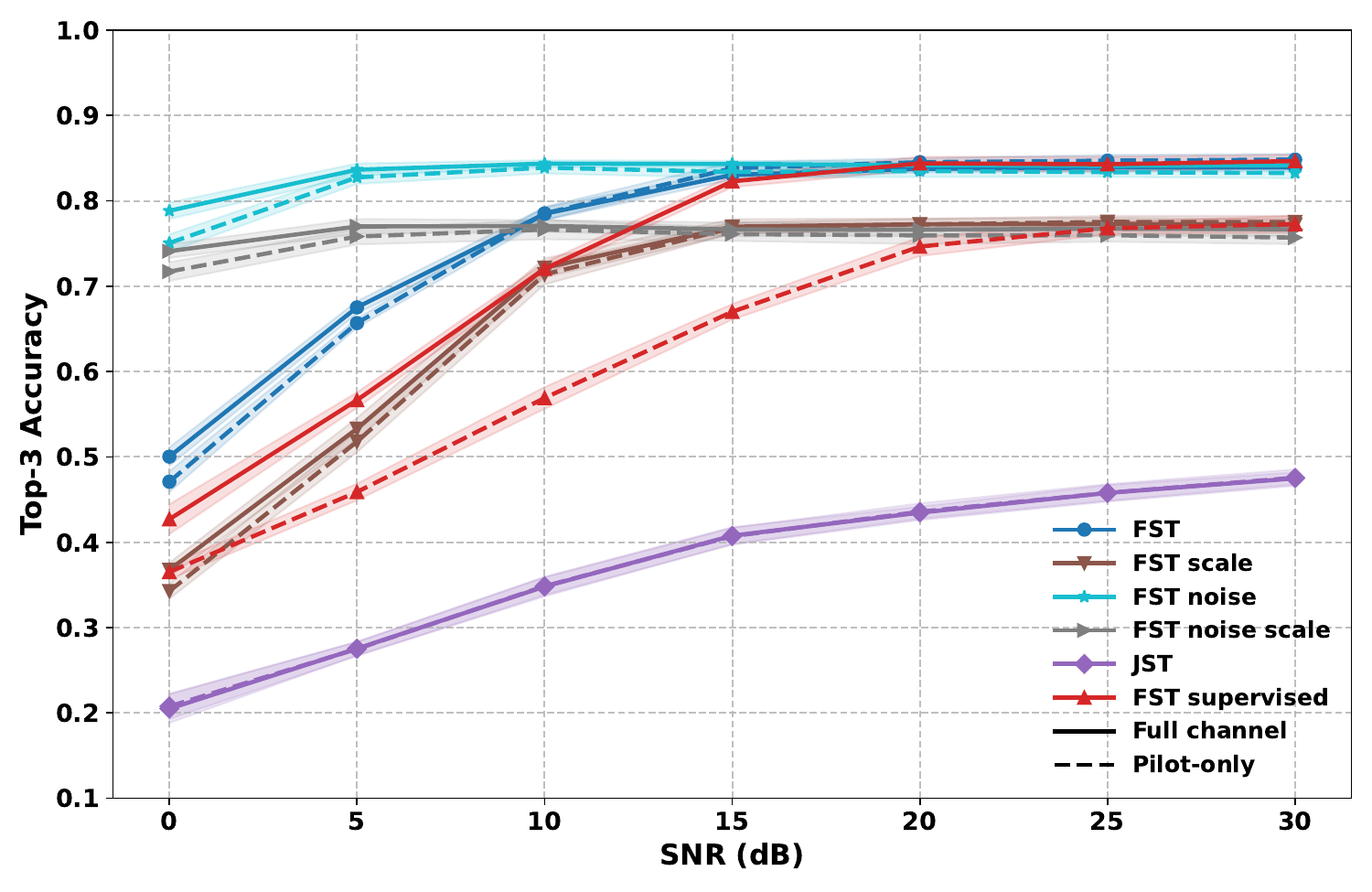}
  \caption{Cross-frequency beam selection (top-3 accuracy) out-of-distribution at 28\,GHz with codebook size $M=128$.}
  \label{fig:beam-ood-28}
\end{figure}

Figs.~\ref{fig:beam-id-28} and~\ref{fig:beam-ood-28} fix $M{=}128$ and compare PilotWiMAE with the supervised baseline, the self-supervised JST encoder, and the intermediate FST ablations in the ID and OOD splits. Several trends emerge. First, low SNR is the differentiating regime. The noise-pretrained variants (FST+noise and FST+noise+scale) dominate at low SNR by a wide margin. This isolates noise-robust pretraining as the dominant low-SNR enabler, rather than the encoder architecture or the supervision signal alone. Second, the supervised baseline exhibits a clear pilot-versus-full gap that opens at moderate SNR and persists across the sweep. Every FST configuration shows curves that nearly coincide between full-channel and pilot-only readouts. This contrast is the empirical signature of pilot-native pretraining. Because the supervised baseline shares the same encoder backbone as PilotWiMAE, the tight pilot-versus-full agreement reflects a property of the pretraining objective, not the encoder. Third, JST trails every FST variant across the entire SNR range and on both splits, supporting the case for the factorized inductive bias under aggressive pretraining masking. Fourth, adding the scale loss on top of noise is mildly negative for beam selection across the SNR range. This is consistent with the joint takeaway that the scale objective benefits channel characterization more than beam selection. The ID-to-OOD shift causes only a small absolute drop and preserves the relative ranking across methods, indicating that cross-band transfer to the held-out city does not introduce method-dependent collapse.

\subsubsection{Channel characterization}

Figs.~\ref{fig:los-id-28} and~\ref{fig:los-ood-28} report the LoS classification accuracy versus SNR on the ID and OOD splits, with the same set of methods as the beam-selection plots. First, the auxiliary scale loss is the dominant ablation, mirroring the beam-selection picture in reverse. The scale-pretrained variants (FST scale and FST noise scale) lead at every SNR and on both splits, since the LoS/NLoS label depends on large-scale fading statistics rather than on fine angular structure. PilotWiMAE (FST noise scale) leads with the flattest profile, since the AWGN curriculum lifts the low-SNR floor. Second, the supervised baseline shows a much smaller pilot-versus-full gap compared to beam selection. This is because recovering a label that depends on aggregate channel power
from sparse pilots is as good as that of the dense grid. Third, the ID-to-OOD shift is essentially invisible, i.e, all curves and their relative ranking are nearly the same for Los Angeles.

\subsubsection{Channel estimation}

\begin{figure}[t]
  \centering
  \includegraphics[width=\linewidth]{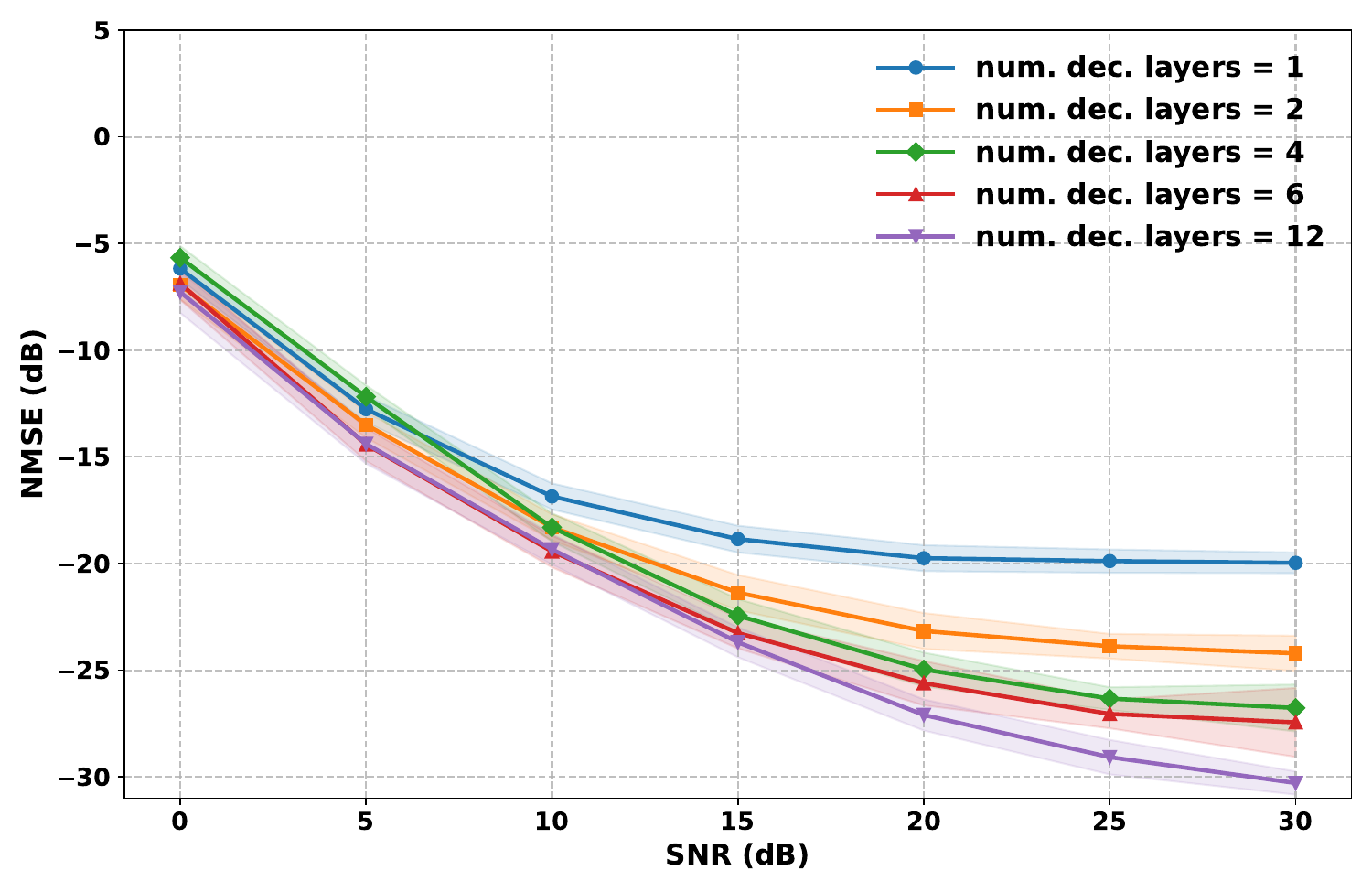}
  \caption{Los Angeles OOD channel estimation at $3.5$\,GHz using a frozen FST+noise+scale encoder with decoder-only pretraining. Curves show NMSE versus SNR for decoder depths 1, 2, 4, 6, and 12.}
  \label{fig:ce-ood-decoder-depth}
\end{figure}

We evaluate dense-channel recovery in the Los Angeles OOD scenario at $3.5$\,GHz for decoder depths $\{1,2,4,6,12\}$. As depicted in Fig.~\ref{fig:ce-ood-decoder-depth}, with increasing depth, NMSE consistently improves throughout the entire range of SNR. The gains are modest at low SNR, where noise dominates, and become more pronounced at medium-to-high SNR. This indicates that additional decoder capacity is most useful once pilot observations are sufficiently reliable.

\begin{figure}[t]
  \centering
  \includegraphics[width=\linewidth]{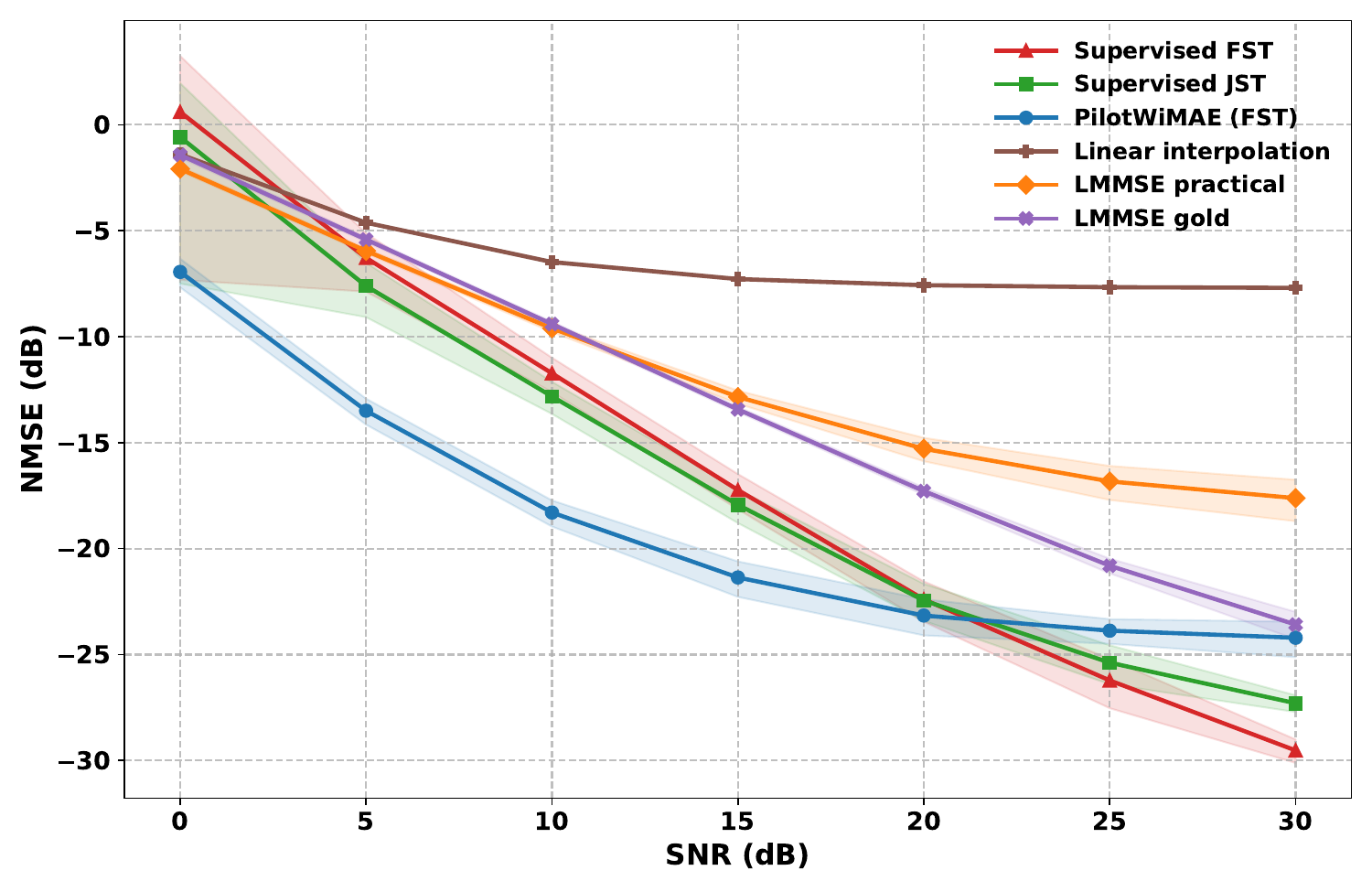}
  \caption{Los Angeles OOD channel estimation at $3.5$\,GHz: NMSE versus SNR for PilotWiMAE (FST), supervised encoder-decoder baselines on the fixed pilot pattern, classical interpolation, and Kronecker LMMSE references (\emph{practical} versus \emph{gold}).}
  \label{fig:ce-ood-comparative}
\end{figure}

Fig.~\ref{fig:ce-ood-comparative} compares the same dense-channel recovery setting against the classical and supervised neural baselines introduced in Section~\ref{sec:experiments:tasks}. PilotWiMAE dominates the low- to mid-SNR regime by a wide margin. At $0$\,dB SNR, PilotWiMAE already reaches an NMSE that the next-best method only attains around $5$\,dB. The lead persists up to around $20$\,dB, where the supervised pilot-pattern-specific baselines catch up and outperform PilotWiMAE at higher SNR. Note that PilotWiMAE's decoder operates on a frozen, pilot-pattern-agnostic encoder representation. On the other hand, the supervised baselines optimize the full encoder-decoder end-to-end against a fixed pilot pattern, which provides more leverage once pilot observations are nearly noise-free. However, the decoder-depth sweep in Fig.~\ref{fig:ce-ood-decoder-depth} shows that scaling the phase-2 capacity to $L_{\mathrm{JST}}=12$ closes the high-SNR gap, after which PilotWiMAE outperforms supervised baselines throughout the entire swept SNR range. Overall, PilotWiMAE is the strongest dense-channel recovery method in the deployment-relevant SNR range.

\section{Computational Complexity}
\label{sec:complexity}

This section reports the training and inference cost of PilotWiMAE and the supervised baseline, profiled with the protocol implemented in our profiler. All measurements use a single NVIDIA RTX A4000 GPU with PyTorch and mixed precision. Training cost is measured at training batch size $B_{\mathrm{tr}}=256$ and is reported in Table~\ref{tab:training-complexity}. Inference cost is measured at inference batch size $B_{\mathrm{inf}}=32$, averaged over $100$ time repeats after warmup, reported in Table~\ref{tab:inference-complexity} as amortized per-sample latency; the amortization convention is described in Appendix~\ref{app:profiling}.
\begin{table}
  \centering
  \caption{Per-epoch training cost on a single NVIDIA RTX A4000 at training batch size $B_{\mathrm{tr}}=256$. FLOPs are reported in petaFLOPs (P).}
  \label{tab:training-complexity}
  \footnotesize
  \setlength{\tabcolsep}{3pt}
  \begin{tabular}{llrrr}
    \toprule
    Method & Encoder & Trainable params & FLOPs/ep. & Time/ep. (s) \\
    \midrule
    PilotWiMAE & FST & $1{,}594{,}658$ & $2.017$ P & $208.01$ \\
    PilotWiMAE & JST & $1{,}594{,}658$ & $2.115$ P & $204.43$ \\
    Supervised baseline           & FST & $1{,}210{,}369$ & $2.656$ P & $653.69$ \\
    \bottomrule
  \end{tabular}
\end{table}

\paragraph{Discussion}
Several observations follow from these tables. First, self-supervised pretraining of either the FST or JST encoder is roughly three times cheaper per epoch than end-to-end supervised training of the same FST backbone. This is because pretraining processes only the visible token subset while the supervised baseline always sees the full grid. Second, in inference, the FST encoder benefits decisively from pilot-only input. Its per-sample latency drops from $0.70$\,ms in the full-channel case to $0.15$\,ms in the pilot-only case, a reduction of $4.6$ times that is consistent with the reduced sequence length combined with the $\mathcal{O}(n_{\mathrm{t}}N_{\mathrm{sf}}(n_{\mathrm{t}}+N_{\mathrm{sf}})d)$ scaling of factorized attention. Third, the JST encoder, whose attention scales as $\mathcal{O}((n_{\mathrm{t}}N_{\mathrm{sf}})^2 d)$ and is therefore asymptotically more expensive than FST's by a factor of $n_{\mathrm{t}}N_{\mathrm{sf}}/(n_{\mathrm{t}}+N_{\mathrm{sf}})$, is nevertheless faster than FST in the pilot-only regime. This is because after masking, its sequence length is small enough that the quadratic cost is no longer the bottleneck and the simpler block structure dominates. However, JST is the slowest of all configurations on the full grid ($1.83$\,ms per sample), where its quadratic dependence on $n_{\mathrm{t}}N_{\mathrm{sf}}$ is exposed.

\begin{table}[t]
  \centering
  \caption{Inference cost on a single NVIDIA RTX A4000 at inference batch size $B_{\mathrm{inf}}=32$, averaged over $100$ timed repeats after warmup.}
  \label{tab:inference-complexity}
  \footnotesize
  \begin{tabular}{llrr}
    \toprule
    Encoder & Input & Encoder params & Per-sample lat.\ (ms) \\
    \midrule
    FST & Pilot only   & $1{,}594{,}658$ & $0.153 \pm 0.007$ \\
    JST & Pilot only   & $1{,}594{,}658$ & $0.073 \pm 0.000$ \\
    FST & Full channel & $1{,}594{,}658$ & $0.698 \pm 0.007$ \\
    JST & Full channel & $1{,}594{,}658$ & $1.833 \pm 0.015$ \\
    \bottomrule
  \end{tabular}
\end{table}

\FloatBarrier
\section{Conclusion}
\label{sec:conclusion}

PilotWiMAE demonstrates that self-supervised wireless channel representation learning can be both robust and deployment-aware by design. Pilot-native inputs avoid unrealistic full-CSI assumptions and factorized attention improves transfer under frequency shift, including to a held-out city, while reducing inference burden. With self-supervised pretraining at 3.5\,GHz, the learned representations transfer to 28\,GHz for beam selection and LoS/NLoS classification without task-specific fine-tuning and provide strong performance under both ID and OOD evaluations. The ablations show that noise-robust pretraining is key for low-SNR stability, and that auxiliary scale supervision is particularly useful for channel-state semantics. These results suggest a practical recipe for future wireless representation learning through structure-aware encoders with deployment-matched corruption and physics-aware pretraining objectives. Extending this recipe to broader pilot patterns, frequencies, and system-level latency profiling is a natural next step. We release the PilotWiMAE pretrained weights and training pipeline, together with the CSIGen ray-tracing channel-generation tool and the channel datasets used in this work, to support reproducibility and future advances in self-supervised wireless channel representation learning.

\FloatBarrier

\appendices

\section{Dataset details}
\label{app:dataset}

For each city, Table~\ref{tab:dataset-city-counts} reports the size of the scene, the number of channels in each split, and LoS prevalence of the test channels. The pretraining cities supply the train split and serve as the ID test set, while the held-out Los Angeles scene supplies the OOD test set. We use $10\%$ of the training set to validate the model during pretraining.

\begin{table}[H]
  \centering
  \caption{Per-city scene size, channel counts, and LoS share.}
  \label{tab:dataset-city-counts}
  \footnotesize
  \setlength{\tabcolsep}{4pt}
  \begin{tabular}{@{}lcrrr@{}}
    \toprule
    City          & Scene size (m)   & Train   & Test   & Test LoS (\%) \\
    \midrule
    Boston        & $697 \times 675$ & 152,675 & 18,423 & 34.07 \\
    New York City & $568 \times 516$ &  92,153 &  8,130 & 27.82 \\
    San Francisco & $605 \times 576$ &  94,262 & 17,755 & 29.84 \\
    Chicago       & $573 \times 567$ &  99,344 & 10,158 & 25.58 \\
    Los Angeles   & $840 \times 754$ &     n/a & 17,466 & 36.04 \\
    \midrule
    Total         & ---              & 438,434 & 71,932 & 31.60 \\
    \bottomrule
  \end{tabular}
\end{table}

\section{Inference pilot pattern}
\label{app:pilot-pattern}

\begin{figure}[H]
  \centering
  \includegraphics[width=\linewidth]{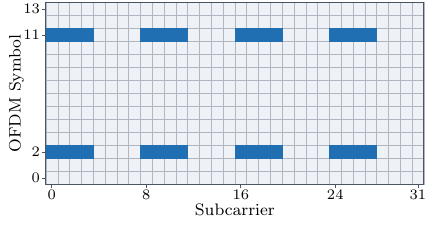}
  \caption{Visualization of the fixed pilot resource elements (highlighted) on the $14\times 32$ OFDM grid.}
  \label{fig:pilot-mask}
\end{figure}

Fig.~\ref{fig:pilot-mask} visualizes the fixed pilot mask.

\section{Per-sample latency convention}
\label{app:profiling}

We report the inference cost per sample, obtained by dividing the per-batch latency by $B_{\mathrm{inf}}$. This amortized convention matches our profiling pipeline and is appropriate as a throughput-style figure of merit in moderate-to-large batches, where the GPU is well-utilized and the per-batch latency scales approximately linearly with $B_{\mathrm{inf}}$. At very small batch sizes, fixed kernel-launch and memory-traffic overhead become non-negligible, so the latency at $B=1$ can exceed the amortized per-sample latency.

\section{Baseline training configurations}
\label{app:baseline-configs}

Table~\ref{tab:jst-train-config} reports the JST pretraining configuration used as the self-supervised baseline. Table~\ref{tab:supervised-train-config} reports the supervised baseline configurations (factorized encoder backbone with a linear classification head).

\begin{table}[H]
  \centering
  \caption{JST pretraining configuration.}
  \label{tab:jst-train-config}
  \footnotesize
  \begin{tabular}{ll}
    \toprule
    Parameter & Value \\
    \midrule
    Encoder type & JST \\
    Input shape & $(14,32,32)$ \\
    Patch shape & $(1,4,4)$ \\
    Embedding & Linear \\
    Positional encoding & Sinusoidal concatenative \\
    Masking & Random, mask ratio $0.95$ \\
    Encoder dimension & $d=128$ \\
    Encoder layers & 6 \\
    Encoder heads & 8 \\
    Decoder layers & 2 \\
    Decoder heads & 4 \\
    Optimizer & AdamW, $\beta=(0.9,0.999)$, wd $0.005$ \\
    Epochs & 500 \\
    Batch size & 512 \\
    LR schedule & Cosine, warmup 10, $\eta_{\min}=10^{-5}$ \\
    Initial LR & $\eta_{\mathrm{start}}=10^{-3}$ \\
    Pos.\ enc.\ scale & $\alpha_{\mathrm{pe}} = 0.01$ at start \\
    Loss / patch norm. & MSE with normalized patch loss \\
    Precision / clipping & Mixed precision, clip at 1.0 \\
    \bottomrule
  \end{tabular}
\end{table}

\begin{table}[H]
  \centering
  \caption{Supervised baseline configurations.}
  \label{tab:supervised-train-config}
  \footnotesize
  \setlength{\tabcolsep}{3pt}
  \begin{tabular}{p{0.26\linewidth}p{0.32\linewidth}p{0.32\linewidth}}
    \toprule
    Parameter & Beam prediction & LoS/NLoS \\
    \midrule
    Encoder backbone & FST & FST \\
    Input shape & $(14,32,32)$ & $(14,32,32)$ \\
    Patch shape & $(1,4,4)$ & $(1,4,4)$ \\
    Encoder dimension & $d=128$ & $d=128$ \\
    Encoder layers & 3 & 3 \\
    Encoder heads & 8 & 8 \\
    Decoder layers & 2 & 2 \\
    Decoder heads & 4 & 4 \\
    Linear head matrix & $d\times M$ & $d\times 2$ \\
    Optimizer & AdamW, wd $0.05$ & AdamW, wd $0.005$ \\
    Epochs & 200 & 200  \\
    Batch size & 256 & 256 \\
    LR schedule & Cosine, warmup 10, $\eta_{\min}=5\!\times\!10^{-6}$ & Cosine, warmup 10, $\eta_{\min}=5\!\times\!10^{-6}$ \\
    Initial LR & $5\times10^{-4}$ & $5\times10^{-4}$ \\
    Loss & Cross-entropy & Cross-entropy \\
    Precision / clipping & Mixed prec., clip 1.0 & Mixed prec., clip 1.0 \\
    \bottomrule
  \end{tabular}
\end{table}

\FloatBarrier

\bibliographystyle{IEEEtran}
\bibliography{references}

\end{document}